\documentclass[a4paper,11pt]{article}
\pdfoutput=1

\usepackage{jinstpub}

\usepackage{lineno}
\usepackage{epstopdf}
\title{Measurement Techniques for Low Emittance Tuning and Beam Dynamics at CESR}

\author{M.G. Billing, J.A. Dobbins, M.J. Forster, D.L. Kreinick, R.E. Meller, D.P. Peterson, G.A. Ramirez,
M.C. Rendina, N.T. Rider, D.C. Sagan, J. Shanks, J.P. Sikora, M.G.
Stedinger, C.R. Strohman, H.A. Williams,\ Cornell Laboratory for
Accelerator-based ScienceS and Education (CLASSE),\ Cornell
University,\ 161 Synchrotron Dr., Ithaca, NY, 14850, U.S.A.\\}
\author{M.A. Palmer,\
Collider Accelerator Department, Brookhaven National Laboratory\
P.O.Box 5000, Upton, NY 11973-5000, U.S.A.\\}
\author{R.L. Holtzapple,\
Physics Department, California Polytechnic State University,\ San
Luis Obispo, CA 93407, U.S.A.\\}
\author{J. Flanagan,\
High Energy Accelerator Research Organization (KEK),\ 1-1 Oho,
Tsukuba, Ibaraki Prefecture 305-0801, Japan}

\abstract{After operating as a High Energy Physics electron-positron
collider, the Cornell Electron-positron Storage Ring (CESR) has been
converted to become a dedicated synchrotron light source for the
Cornell High Energy Synchrotron Source (CHESS).  Over the course of
several years CESR was adapted for accelerator physics research as a
test accelerator, capable of studying topics relevant to future
damping rings, colliders and light sources.  Initially some specific
topics were targeted for accelerator physic research with the
storage ring in this mode, labeled CesrTA.  These topics included
1)~tuning techniques to produce low emittance beams, 2)~the study of
electron cloud (EC) development in a storage ring and 3)~intra-beam
scattering effects.  The complete conversion of CESR to
CesrTA~occurred over a several year period, described
elsewhere\cite{JINST10:P07012, JINST10:P07013,JINST11:P04025}. A
number of specific instruments were developed for CesrTA.  Much of
the pre-existing instrumentation was modified to accommodate the
scope of these studies and these are described in a companion
paper\cite{JINST12:T11006}. To complete this research, a number of
procedures were developed or modified, often requiring coordinated
measurements among different instruments\cite{CLNS:12:2084}.  This
paper provides an overview of types of measurements employed for the
study of beam dynamics during the operation of CesrTA.}


\begin{document}
\maketitle
\flushbottom

\section{Introduction}

The CesrTA project was defined initially with a set of specific
goals for accelerator physics research and development:

\begin{itemize}
\item The investigation of instrumentation and methodology to systematically reduce the vertical emittance of the stored
beam.\cite{CornellU2013:PHD:JShanks, PRSTAB17:044003}
\item The measurement and characterization of the intra-beam scattering effects as they lead to emittance
enlargement.\cite{CornellU2013:PHD:MPErhlichman,PRSTAB16:104401,IPAC14:TUPRI035}
\item The study of the effects on the stored positron beam due to electron clouds (EC), produced by synchrotron radiation-induced photoelectrons.
The goal is to 1) characterize and quantify the production
mechanisms, 2) compare these with computer simulations, 3) develop
methods to mitigate the EC effects, and 4) study the related beam
dynamics and instabilities.\cite{CornellU2013:PHD:JCalvey,
PRSTAB17:061001, NIMA760:86to97, NIMA770:141to154, PRAB19:074401}
\end{itemize}

To meet these goals a number of instruments in CESR required
significant upgrades or complete development  of new designs and
subsequent implementation.  The simultaneous or coordinated data
acquisition of the ensemble of instruments, which is required during
specific measurement sequences is described below in the following
broad categories:

\begin{itemize}
\item Beam-Based Measurement Techniques
\item Performance of Low Emittance Instrumentation
\item Coordinated Dynamics Measurements
\end{itemize}

The following sections include descriptions of the measurements that
are undertaken and some of the analyses that are performed to meet
the goals.  Note that these sections are not intended to present
detailed accelerator physics measurements as this is reserved for
papers, which can present the motivation and the experimental
details for the different sets of observations.

\section[Overview of the Requirements for Beam Instrumentation]{Overview of the Requirements for Beam Instrumentation}

The particle beam instrumentation developed specifically for
CesrTA~or adapted from CESR's existing instrumentation has been
employed to study low emittance beams, intra-beam scattering and EC
beam dynamics.  This section provides an overview for some of the
fundamental measurements that have been undertaken in each of these
studies.

\subsection{Low Emittance Tuning Measurements}
\label{ssec:lo_emit_meas}

Tuning methods have been developed, which has allowed a significant
reduction of the ratio of the vertical emittance with respect to the
horizontal emittance.  This method is described in detail in these
references\cite{PRSTAB17:044003, CornellU2013:PHD:JShanks}. After
performing beam position monitor (BPM) gain corrections and
centering, the procedure to reduce vertical emittance uses the CESR
beam position monitor (CBPM) system\cite{JINST12:T09005} in a
sequence of measuring and correcting 1)~the closed orbit,
2)~betatron phase and coupling, and 3)~vertical dispersion. Using
the x-ray beam size monitor (xBSM)\cite{NIMA798:127to134}, the
success of this procedure is determined by measuring the vertical
beam size at each step of the
process\cite{CornellU2013:PHD:JShanks}.

\subsection{Intra-Beam Scattering Measurements}
\label{ssec:ibs_meas}

Studies of intra-beam scattering have been undertaken at
CesrTA~\cite{PRSTAB17:044002, PRSTAB16:104401, IPAC13:TUODB103,
IPAC13:TUPME065, IPAC12:WEPPR015}.  The operating tunes for the
storage ring were measured using either 1)~the set of shakers to
drive the beam and then observe a stretched BPM signal on a spectrum
analyzer or 2)~the tune tracker and turn-by-turn CBPM measurements,
which are then Fourier transformed.   At the selected tunes and
after performing the low emittance tuning procedure described in
section~\ref{ssec:lo_emit_meas}, a set of observations were made as
a function of the number of particles in a single bunch.  The
instruments utilized to record these observations are 1)~the xBSM to
measure the vertical beam size turn-by-turn, 2)~the visible-spectrum
beam size monitor (vBSM) to measure the horizontal beam size and
3)~a streak camera to measure the bunch length via the visible light
port.  The data is typically recorded every few seconds as the beam
current decays.

\subsection{Electron Cloud Dynamics Measurements}
\label{ssec:ec_dyn_meas}

Photo-electrons produced from synchrotron radiation photons
impinging on the vacuum chamber wall are accelerated by stored
positron bunches.  If several bunches are closely spaced to form a
train, then the accelerated electrons impact the vacuum chamber,
producing secondary electrons, which are again accelerated.  If the
surface of the wall has a secondary emission yield larger than one,
then the number of electrons in the EC multiply. This cloud can
react back on the beam affecting its dynamics. At CesrTA~ dynamics
studies of the EC's interaction with the beam fall into three broad
classes of measurements, 1)~the tune shift due to the EC through a
train of bunches, 2)~the beam position spectra and vertical beam
sizes of bunches within the train and 3)~coherent damping of dipole
and head-tail transverse modes vs. the position of bunches within a
train\cite{PAC11:WEP194, ECLOUD10:DYN03, ECLOUD10:PST02}.

Tune shift measurements taken for bunches within the train allow the
estimation of the EC density in the neighborhood of the bunches'
orbit.  The vast majority of the measurements of tune shift have
been taken in two different configurations.  In the first horizontal
and vertical pinger magnets\cite{JINST12:T11006} are independently
excited to deflect all of the bunches, causing them to execute
betatron oscillations in the respective transverse plane.  CBPM
turn-by-turn data, triggered just before the pingers fire, is
acquired for all of the bunches within the train. This process may
be repeated as the current in the bunches is uniformly increased.
The CBPM data are analyzed offline and the two tunes are identified
for each bunch. This method is relatively fast to acquire, but
conclusions from these measurements must be tempered as later
bunches in the train are actually oscillating in response to the
excitation from the moving EC, which itself is being driven at
frequencies present from preceding bunches.  In CesrTA~ this effect
is especially pronounced in the horizontal plane.  The second method
uses horizontal and vertical stripline kickers to drive the lead
bunch in the train and a second bunch, positioned later in the
train.  Again CBPM data are taken turn-by-turn for all bunches
within the train and the data are Fourier transformed to produce the
betatron spectra of the bunches. This method allows an accurate
measurement of the tune shift of the later bunch with respect to the
lead bunch without driving the EC into an oscillation prior to the
arrival of the later bunch.  Unlike the first method, this requires
taking many sets of data to span the length of the train, but it
more accurately reflects the tune shift associated with a ``static"
EC\cite{IPAC11:MOPS084}.

Characterizing the instabilities of the bunches within a train due
to the EC is the motivation for the second class of measurements.  A
train of bunches is filled to a uniform current per bunch, at which
unstable motion has been detected.  Then the gated signal from a
single BPM button is detected for a single bunch as is described
below.  After processing by a spectrum analyzer, the spectrum is
recorded bunch-by-bunch.  While this process continues to step
through the bunches in the train, the xBSM is triggered to measure
the beam size of every bunch turn-by-turn.  By suitable choices for
the settings of the beam stabilizing feedback and the chromatic
damping, spectral lines at the head-tail frequencies may be
observed. From this data the frequency and equilibrium amplitude of
the self-excited bunches and the growth of their vertical beam sizes
along the train may be determined\cite{IPAC13:TUPWA063,
IPAC12:WEPPR087, IPAC13:TUPWA061, IPAC13:TUPWA062, IPAC12:WEYA02,
IPAC11:MOPS084}.

The third class of measurements has the same hardware configuration
as the preceding class of spectrum measurements.  It has been
developed to study the damping of coherent betatron motion of the
bunches within train to understand any mechanisms, by which the EC
cause the motion to damp or de-cohere.   This is accomplished by
using a stripline kicker, driving one of the transverse dipole or
head-tail modes, to excite the motion of a single bunch for a short
length of time (e.g. 1~msec) and then observing the damping of the
oscillation using a spectrum analyzer as a tuned receiver at the
exciting betatron tune frequency.  This measurement provides
information about the imaginary part of the tune shift due to the
EC\cite{IPAC11:MOPS084}.  Another variation of drive-damp
measurements excites one or both dipole transverse modes of one
bunch within a train for a short duration while recording the
turn-by-turn positions of all bunches using the CBPM system.
Although this second method is only sensitive to the dipole motion
of the bunches, it allows the observation of how the motion of one
bunch excites subsequent bunches via coupling through the EC.



\section[Beam-based Measurement Techniques]{Beam-based Measurement Techniques}
\label{sec:let.beam_measurements} The low emittance tuning procedure
relies on beam based measurements both to calibrate the beam
position monitors and to determine sources of emittance dilution.
The tuning procedure typically begins with a measurement of the
closed orbit. As described in \cite{JINST12:T09005}  the beam
position monitors have a single shot resolution of about 10 $\mu$m.
After correcting closed orbit errors, the betatron phase and
coupling are measured and corrected as described below. Subsequently
the same measurements and corrections are performed for the vertical
dispersion. Techniques for determining the linear lattice functions
are described later in this section.

Systematic effects limit the accuracy of our measurements of closed
orbit, vertical dispersion and, thus, transverse coupling.
The most significant systematic error in the measurement
of the closed orbit is a result of the uncertainty of the offset of the BPM with respect to
the center of the adjacent quadrupole. Beam based BPM/quad centering
is discussed below. The accuracy of the measurement of vertical
dispersion is limited by variations in the gains of the BPM button
electrodes and related electronics and the physical tilt of the beam
position monitors. We attempt to resolve these systematics with beam based
measurement techniques.

\subsection{Phase and Coupling Correction}
\label{ssec:let.beam_instr.phase_coupling}

Betatron amplitude, phase and transverse coupling are measured
simultaneously. A circulating bunch is driven resonantly with
magnetic shakers or stripline kickers by means of tune trackers at
the transverse normal mode tunes. Details of this method are
described in \cite{PRSTAB3:092801, PRSTAB3:102801}, and are briefly
summarized here.

\begin{figure}[tb] 
   \centering
   \includegraphics[width=2.25in]{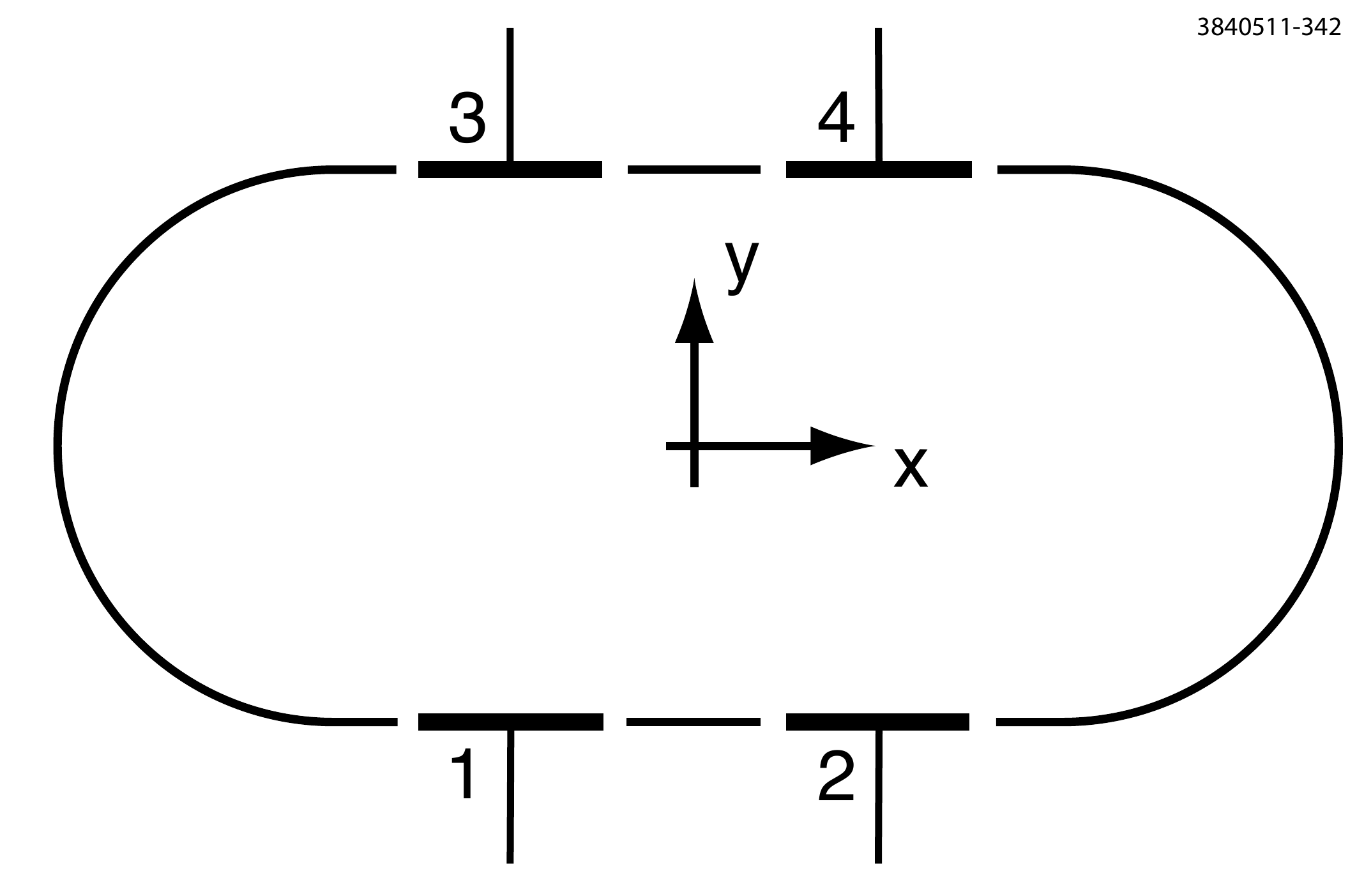}
  \caption[Button layout at a beam position detector.]{\label{fig:let:bpm} Button layout at a beam position detector}
\end{figure}

At each of the four button electrodes (see Figure~\ref{fig:let:bpm})
of each beam position monitor, the intensity of the oscillation
signal is measured turn-by-turn. Typically $N=40,000$ turns of data
are recorded in the CBPM module. For the motion at the
``horizontal'' frequency the amplitude and phase of the motion at
each button is extracted as follows

\begin{eqnarray}
A_{j,sin,h} &=& {2\over N} \sum_{i=1}^N \sin(\theta_{t,h}(i)) \, a_j(i) \nonumber \\
A_{j,cos,h} &=& {2\over N} \sum_{i=1}^N \cos(\theta_{t,h}(i)) \,
a_j(i)
  \label{asinacos}
\end{eqnarray}

\noindent where $A_{j,sin,h}$ and $A_{j,cos,h}$ are the ``in-phase''
and ``out-of-phase'' components of the horizontal beam motion at
button $j = 1, \ldots, 4$. In the above equation $\theta_{t,h}(i)$
is the phase of the reference signal from the horizontal tune
tracker at turn $i$, which is encoded by the timing system onto the
turn-by-turn CBPM trigger, and $a_j(i)$ is the signal on button $j$
at turn $i$. There is a similar equation for the vertical motion. In
order to speed up the data taking process, horizontal and vertical
measurements are done simultaneously. The above analysis is
performed within the digital signal processor (DSP) of each CBPM
module and depends upon the tunes being well enough separated from
any low-order resonance so that the ``cross-talk'' between the
horizontal and vertical mode sums in Eqs.~\ref{asinacos} is
negligible. In practice it is always easy to satisfy this condition.

After the button amplitudes have been measured and summed, the in-phase
amplitudes in the $x$ and $y$ planes from the horizontal excitation are given by

\begin{eqnarray}
A_{x,sin,h} &=& C_x \, \frac{A_{4,sin,h} + A_{2,sin,h} - A_{3,sin,h}
- A_{1,sin,h}}
                           {A_{4,sin,h} + A_{3,sin,h} + A_{2,sin,h} + A_{1,sin,h}} \nonumber \\
A_{y,sin,h} &=& C_y \, \frac{A_{4,sin,h} + A_{3,sin,h} - A_{2,sin,h}
- A_{1,sin,h}}
                           {A_{4,sin,h} + A_{3,sin,h} + A_{2,sin,h} + A_{1,sin,h}}
  \label{axay}
  \end{eqnarray}

Similar equations describe the out-of-phase component at the
horizontal tune and then a corresponding set gives the in-phase and
out-of-phase response to vertical shaking. $C_x$ and $C_y$ are
spatial constants dependent upon the geometry of the BPM. In the
equations above the dependence of position on button amplitude is
illustrated assuming linear dependence of the BPM response. In
practice we include nonlinear dependence utilizing a nonlinear model
of the BPM pickup response\cite{PRSTAB8:062802}.

 The next step is to turn the sine and cosine components into
phase and amplitude
\begin{eqnarray}
A_{x,h} &=&\sqrt{(A_{x,sin,h})^2 + (A_{x,cos,h})^2} \nonumber \\
\phi_{x,h} &=& \tan^{-1}{A_{x,sin,h}\over A_{x,cos,h}}
 \label{axhphixh}
\end{eqnarray}
and similarly for vertical shaking.
In general, the horizontal mode motion of the beam can be written
in the form~\cite{PRSTAB2:074001}
  \begin{eqnarray}
x &=&  \gamma_c \, a_x \, \sqrt{\beta_h} \, \cos(\psi_{h,i}) \nonumber \\
y &=&             -a_x \, \sqrt{\beta_v} \, [\overline C_{22} \,
\cos(\psi_{h,i}) +
                                            \overline C_{12} \,
                                            \sin(\psi_{h,i})]
 \label{xyh}
  \end{eqnarray}
where $\beta_h$ and $\beta_v$ are the beta functions at the BPM,
$\overline C_{ij}$ are components of the coupling matrix, and
  \begin{equation}
\psi_{h,i} = i \, \mu_h + \phi_{h,\beta} + \phi_{h,0}
  \label{phmpp}
  \end{equation}
with $i$ being the turn number, $\mu_h$ the horizontal tune,
$\phi_{h,\beta}$ the horizontal betatron phase at the BPM, and
$\phi_0$ an overall phase offset independent of the BPM. In the
above equation $\gamma_c$ is a parameter dependent upon the
coupling.  If the local horizontal-to-vertical coupling is small,
$\gamma_c$ is very nearly unity and can be approximated as such.

Comparison of Eq.~\ref{xyh} with the measured beam amplitude components gives
  \begin{eqnarray}
\beta_h &=& k_{h,\beta} \, A_{x,h}^2 \nonumber \\
\phi_{h,\beta} &=& \phi_{x,h} + d\phi_h \nonumber \\
\overline C_{12} &=&  \sqrt{\frac{\beta_h}{\beta_v}} \,
\frac{A_{y,h}}{A_{x,h}} \, \sin(\phi_{y,h} - \phi_{x,h})
\label{eq:cbar12_y_x} \\  \overline C_{22} &=&
-\sqrt{\frac{\beta_h}{\beta_v}} \, \frac{A_{y,h}}{A_{x,h}} \,
\cos(\phi_{y,h} - \phi_{x,h}) \nonumber
  \end{eqnarray}
where $k_{h,\beta}$ and $d\phi_h$ are overall constants dependent upon
the amplitude and phase of the shaker drive.  A similar analysis is
carried out for the vertical shaking data. Here the $x$ and $y$
motions are
  \begin{eqnarray}
x &=&              a_y \, \sqrt{\beta_h} \, [\overline C_{11} \, \cos(\psi_{v,i}) - \overline C_{12} \, \sin(\psi_{v,i})] \\
y &=&  \gamma_c \, a_y \, \sqrt{\beta_v} \, \cos(\psi_{v,i})
  \end{eqnarray}
where $\psi_{v,i}$ is analogous to $\psi_{h,i}$ in Eq.~\ref{phmpp}
  \begin{equation}
\psi_{v,i} = i \, \mu_v + \phi_{v,\beta} + \phi_{v,0}
  \end{equation}
Comparison of this with the measured beam amplitude components gives
  \begin{eqnarray}
\beta_v &=& k_{v,\beta} \, A_{x,v}^2 \nonumber \\
\phi_{v,\beta} &=& \phi_{x,v} + d\phi_v \nonumber \\
\overline C_{12} &=&  \sqrt{\frac{\beta_v}{\beta_h}} \, \frac{A_{x,v}}{A_{y,v}} \, \sin(\phi_{x,v} - \phi_{y,v}) \\
\overline C_{11} &=& -\sqrt{\frac{\beta_v}{\beta_h}} \,
\frac{A_{x,v}}{A_{y,v}} \, \cos(\phi_{x,v} - \phi_{y,v}) \nonumber
  \end{eqnarray}
The data is collected and analyzed for all 100 BPMs in less than 10 seconds.

\begin{figure}[tb] 
   \centering
   \includegraphics[width=0.9\textwidth]{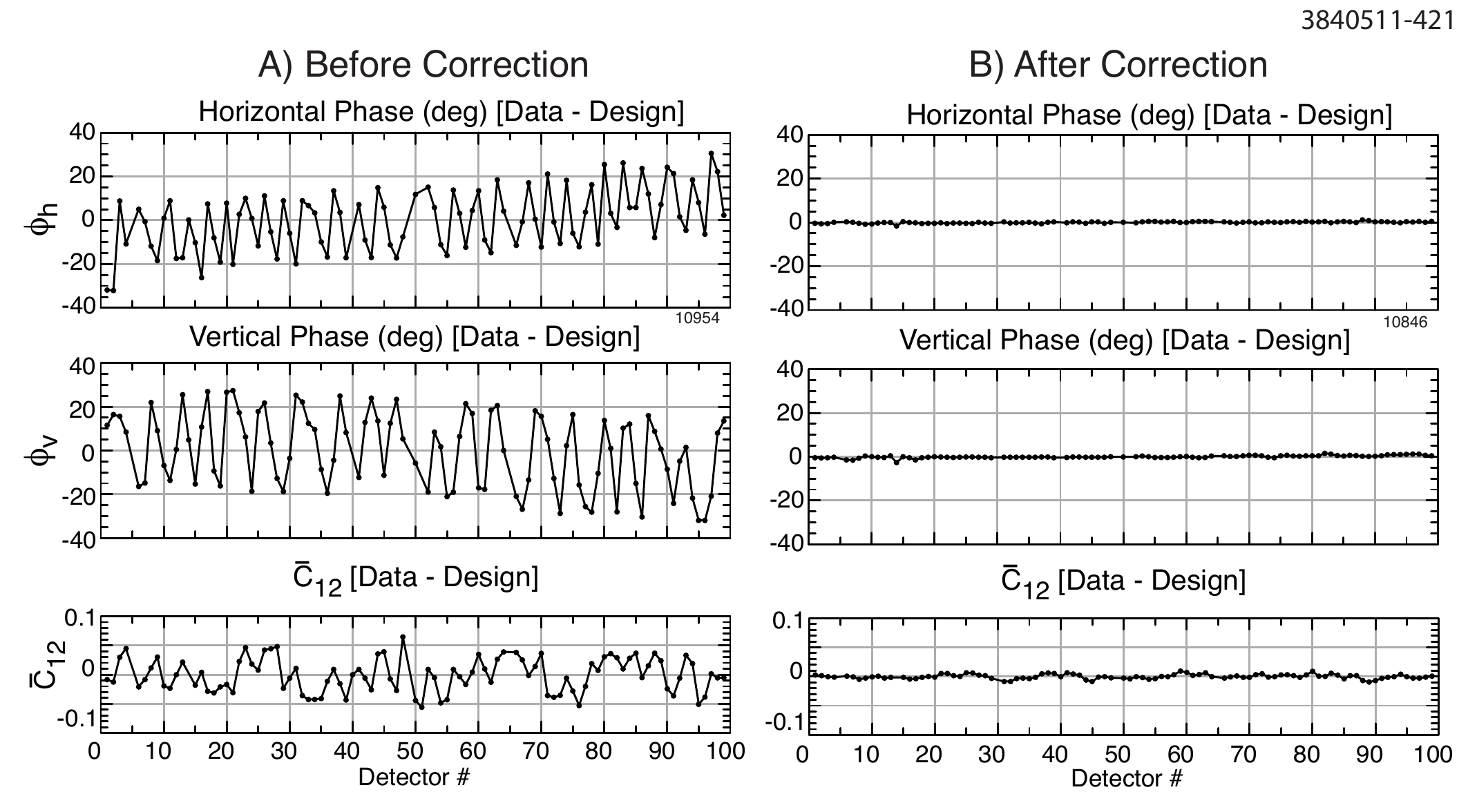}
  \caption[Example betatron phase and coupling correction.]{\label{fig:correct}
Example betatron phase and coupling correction. A) Measured betatron
phases and $\overline C_{12}$ coupling matrix element (as computed
by equation \ref{eq:cbar12_y_x}) before a correction. Plotted are
the difference between the measured values and the design values as
computed from the theoretical design lattice. B) Measured values
after correction. A perfect correction would result in the measured
values being equal to the design ones.}
\end{figure}

Once a measurement is completed, the independently-controlled strengths of the quadrupoles and skew
quadrupoles can be adjusted to correct the Twiss parameters and
coupling. This is done using a software model of the CESR
lattice. The correction procedure occurs in two steps. The first step is
to vary the quadrupoles and skew quadrupoles in the model until the
betatron phases and the $\overline C_{12}$ as computed from the model
most closely match the measured phases and measured $\overline
C_{12}$. The measured beta is generally ignored since the
betatron phase gives a more accurate signal. This is true since the betatron
phase is insensitive to variations in the gains of the button
signals. Similarly, the $\overline C_{11}$ and $\overline C_{22}$ data
is typically ignored since the $\overline C_{12}$ data is generally more accurate.
This is due to the insensitivity of the $\overline C_{12}$
data to button gain errors and rotations of the BPM in the $x$-$y$ plane.

An example correction is shown in Figure~\ref{fig:correct}.
Figure~\ref{fig:correct}A shows the difference between measured and
design betatron phase and coupling before correction. Since
the betatron phase can always be offset by an arbitrary additive
constant, the phase constants $d\phi_h$ and $d\phi_v$ are taken to be
such that the average measured phase is equal to the average phase as
calculated from the design lattice. Figure~\ref{fig:correct}B shows the
measured phase and coupling after correction. The deviation between actual and design
values is now quite small. The vertical emittance ratio
$\epsilon_v/\epsilon_h$ due to coupling effects is roughly
$\sigma^2_{\overline C_{12}}$, the square of the root mean squared (RMS) standard deviation of
$\overline C_{12}$. After correction, $\sigma_{\overline C_{12}}$ is
0.004 so $\epsilon_v/\epsilon_h$ due to coupling is of order $2 \times
10^{-5}$, which is very small. The RMS betatron phase deviation after
correction is about 0.5~degrees, which translates to an RMS deviation
of the actual $\beta$ relative to the design $\beta$ of less than 1\%.

\subsection{Dispersion}
\label{ssec:let.beam_instr.dc_dispersion}

Dispersion is measured via the change in the closed orbit for a
given change in beam energy. The beam energy is related to the RF
frequency via the momentum compaction, typically $\alpha_p=6.8\times
10^{-3}$ for low emittance CesrTA optics. RF cavity bandwidth and
energy aperture limit the frequency change to $\pm4$ kHz of the
500~MHz accelerator cavity frequency and energy offset to about
0.24\%. Sensitivity to a vertical dispersion of 1 cm, requires that
we measure an orbit difference of $24\,\mu$m. The CesrTA beam
position monitors nominally have turn-by-turn reproducibility of
order $10\,\mu$m and the orbit measurement is based on an average of
several thousand turns.

\subsection{BPM/quad centering}
\label{ssec:let.beam_instr.quad_centering}

To reach the lowest vertical emittance it is necessary for the
vertical closed orbit to be as close to the centers of the quadrupoles
as possible to minimize vertical kicks, which would contribute to the
vertical dispersion. Correction of the closed orbit can be compromised
by offsets between the magnetic center of a quadrupole and the
electrical center of a nearby BPM.

The standard beam based technique to determine these offsets involves the
measurement of orbit changes as the quadrupole's
strength $k$ is varied\cite{EPAC04:MOOCH01}. One common method for determining
the quadrupole center involves taking measurements at various beam
positions and then interpolating the results to find the quadrupole
center, which is the position where the beam orbit does not change with
variation of the quadrupole strength. This method can be slow due to
the number of measurements needed. For large rings, this technique is
not practical.

The measurement method developed at Cornell uses a variation of this
procedure:
  \begin{enumerate}
  \item
Betatron phases and orbits are measured at two
different strength settings for a given quadrupole at some location
$s_q$. To differentiate the two measurements, one measurement is
called the ``base'' measurement and is denoted with a subscript ``0''
and the other is called the ``non-base'' measurement.
  \item
The change in vertical orbit $dy \equiv y - y_0$ at the location s, due to the variation in quadrupole
strength, is given by
\begin{equation}
  dy(s) = dy' \,
  \frac{\sqrt{\beta_y(s) \, \beta_y(s_q)}}{2 \, \sin \pi\nu_y}
  \, \cos (|\phi_y(s) - \phi_y(s_q)| - \pi \nu_y)
  \label{eq:dydy}
\end{equation}
where $dy'$ is the kick due to the change in quadrupole strength, and
the Twiss parameters $\beta$, $\phi$ and $\nu$ are evaluated in the
non-base configuration. From the phase data, the Twiss parameters can be
calculated. Using the calculated Twiss parameters, a fit to the orbit difference data
gives a measurement of $dy'$.
  \item
From a fit of the betatron phase data, the change in quadrupole strength $dk$
and the beta at the quadrupole can be calculated. The change in
orbit $dy$ can be written in the form
\begin{equation}
  dy(s) = [y_{qc} - y_0(s_q)] \, dk \, L \,
  \frac{\sqrt{\beta_y(s) \, \beta_y(s_q)}}{2 \, \sin \pi\nu_y}
  \, \cos (|\phi_y(s) - \phi_y(s_q)| - \pi \nu_y)
  \label{eq:dyyydk}
\end{equation}
where $y_{qc}$ is the quadrupole center, and $L$ is the quadrupole length.
Comparing the above two equations gives
\begin{equation}
  y_{qc} = \frac{dy'}{L \, dk} + y_0(s_q)
\end{equation}
A similar analysis holds for the horizontal plane.
  \item
The accuracy of the calculation depends upon the displacement of the
beam from the quadrupole center $dy_{beam}$. This separation is given by
\begin{equation}
  dy_{beam} = y_{qc} - y_0(s_q) = \frac{dy'}{L \, dk}
\end{equation}
The closer the beam is to the quadrupole center, the more accurate the calculation of $y_{qc}$.
If the beam is too far off from the center, an orbit bump
is used to steer the beam towards the center and the measurement cycle is
repeated until $|dy_{beam}|$ is within a set tolerance. Typically this is
$300 \, \mu$m.
  \end{enumerate}
The important innovation here is that the analysis incorporates the
phase measurement which leads to a more accurate determination of
the Twiss parameters and $dk$. This leads to a more accurate
determination of $y_{qc}$. The result is that the tolerance on
$dy_{beam}$ can be increased, which leads to a reduced number of
measurement cycles and hence a shorter measurement time. The entire
procedure in CESR --- calibrating all 100 or so BPMs --- has been
automated and takes somewhat less than two hours.


\subsection{BPM Gains}
\label{ssec:let.beam_instr.bpm_gains}

The principal systematics limiting the accuracy of our measurement
of vertical dispersion are variation of BPM button gains and BPM
tilts. There are several effects that will contribute to gain
errors, beginning with the button itself and extending through the
cabling and electronics. Small variations in how the button is
seated in the BPM block will affect the response to the beam signal,
as will cable and connector dependent attenuation. Finally, the four
BPM button electrodes are connected to dedicated front end
amplifiers, each of which can respond with slightly different gains.

\subsubsection{Effect from BPM Gain Errors on Vertical Dispersion}
\label{ssec:let.beam_instr.v_dispersion_from_bpm_gains}

It is easy to see that the measurement of vertical dispersion is
especially sensitive to gain errors. Suppose, for example, at a particular
detector there is finite horizontal, but identically zero vertical
dispersion. If there is any variation in button gains, a component of horizontal
dispersion will appear as vertical dispersion in the measurement.  An estimate of the
sensitivity of the measurement of vertical dispersion can be made from
gain errors with this simple example. Consider a BPM where there is finite horizontal
dispersion $\eta_x$ and zero vertical dispersion, and the on-energy
orbit is
at precisely zero in both horizontal and vertical.
The position for an energy offset, +$\Delta E/2$,  is given by

\begin{eqnarray}
x^+ = x_0\left({B^+_4-B^+_3+B^+_2-B^+_1\over \sum B_i}\right)
\end{eqnarray}

\noindent where $x_0=19.6$ mm is the typical BPM horizontal scale
factor and the labels are defined in Figure~\ref{fig:let:bpm}. We
assumed that there is zero vertical orbit offset, and since there is
zero vertical dispersion and no change in vertical position with the
energy, it must be that $B^+_4=B^+_2$ and $B^+_3=B^+_1$ so that

\begin{eqnarray}
x^+ = 2x_0\left(\frac{B^+_4-B^+_3} {\sum B_i}\right)
\end{eqnarray}

\noindent Then with energy offset $-\Delta E/2$,

\begin{eqnarray}
x^- = x_0\left(\frac{B^-_4-B^-_3+B^-_2-B^-_1}{\sum B_i}\right)
\end{eqnarray}

\noindent By symmetry, $B^-_3 = B^+_4$ and $B^-_1 = B^+_2$.
Therefore

\begin{equation}
\Delta x = x^+-x^-= 4x_0\left(\frac{B^+_4-B^+_3}{\sum
    B_i}\right)\label{deltax}
\end{equation}

\noindent and $\eta_x = \Delta x / \Delta E$. Now suppose that there
is a fractional gain error $f$ for $B_3$ so that $B_3\rightarrow
(1+f)B_3$. Then we will measure

\begin{eqnarray}y^+ = y_0\left({fB^+_3\over \sum B_i}\right),\ \ y^-=
y_0\left({fB^-_3\over \sum B_i}\right)= y_0\left({fB^+_4\over
    \sum B_i}\right)
\end{eqnarray}

\noindent since $B_4^+=B_3^-$. Finally,

\begin{equation}
\Delta y =  y_0\left(\frac{f(B^+_3-B^+_4)}
    {\sum B_i}\right)\label{deltay}
\end{equation}

\noindent ($y_0=26$ mm for a typical CESR BPM.) Combining the
equations ~\ref{deltax} and ~\ref{deltay} we find that

\begin{eqnarray}
\Delta y ={f\over 4}\Delta x {y_0\over x_0}\rightarrow \Delta \eta_y
\sim 0.34 f\eta_x
\end{eqnarray}

\noindent The peak horizontal dispersion in the CesrTA lattice is
about 2 m. Therefore a 1\% gain error on a single button will
introduce an error in the measurement of the vertical dispersion of
$\Delta~\eta_y\sim~6.8$~mm.

\subsubsection{BPM Gain Calibration Procedure}
\label{ssec:let.bpm_gain_calibration}

\begin{figure}[htb] 
   \centering
   \includegraphics[width=5in]{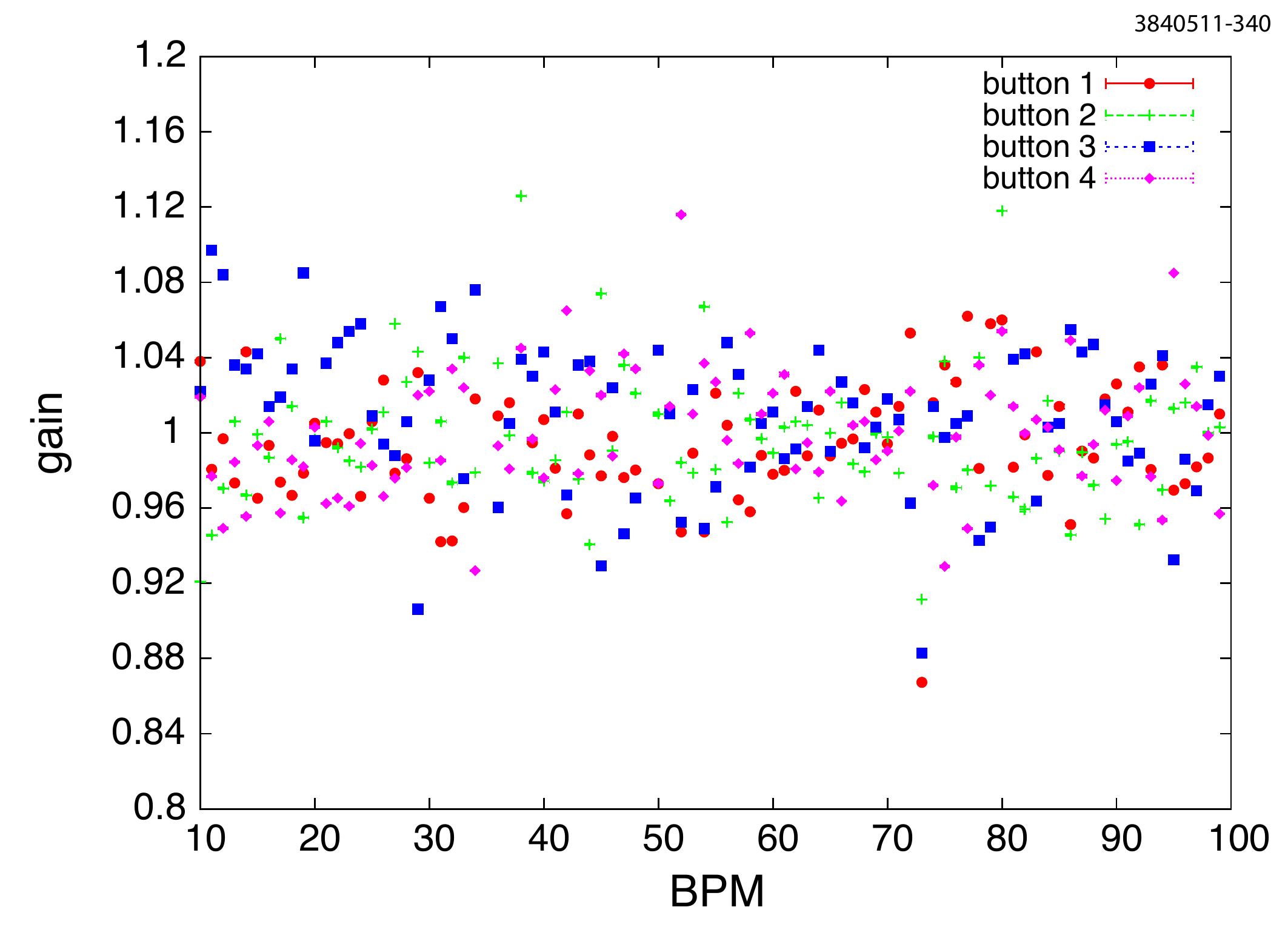}
  \caption{\label{fig:let:all_gains} Measured BPM button gains. Note
    that there are four data points for each BPM.}
\end{figure}

To calibrate the gains, we take advantage of the fact that a position
measurement in a 4 button detector is over constrained. Indeed, the
signals on any
three of the four buttons locates the beam in the detector. Expanding
the button signals to second order in the beam's displacements, x and y, we find
that the four button signals are related as follows\cite{PRSTAB13:092802}
\begin{equation}
B_1^i-B_2^i-B_3^i+B_4^i =  {c\over I}(B_1^i-B_2^i+B_3^i-B_4^i
)(B_1^i+B_2^i-B_3^i-B_4^i)
\end{equation}
where $B_j^i$ is the intensity on the $j^{th}$ button on the $i^{th}$
measurement. $c$ is a constant determined by the BPM geometry and $I$
the bunch current. If we suppose that there is a variable gain
associated with each button then we can write
\begin{multline}
g_1B_1^i-g_2B_2^i-g_3B_3^i+g_4B_4^i =  \\
   {c\over I}(g_1B_1^i-g_2B_2^i+g_3B_3^i-g_4B_4^i)
 (g_1B_1^i+g_2B_2^i-g_3B_3^i-g_4B_4^i)~~~~
\end{multline}
The next step is to measure response to the beam over the entire
active area of the detector so that we have a set $B_j^i$, for
$i=1,\ldots,N$ where $N$ is a large number (typically 1024 turns).
Then we fit the $g_j$ to minimize
\begin{multline}
\chi^2=
\sum_{i=1}^N\left[g_1B_1^i-g_2B_2^i-g_3B_3^i+g_4B_4^i -  {c\over I}(g_1B_1^i-g_2B_2^i+g_3B_3^i-g_4B_4^i)    \right. \\
  \left. (g_1B_1^i+g_2B_2^i-g_3B_3^i-g_4B_4^i)\right]^2~~~~
\end{multline}
In order to sample the active region of all of the BPMs, we resonantly drive the beam simultaneously at the
horizontal and vertical tunes, and collect turn-by-turn data.
Because the measurement and fitting can be completed in just a few
minutes, it is straightforward to maintain up-to-date gain
calibrations. An example of fitted gains for all 100 beam position
monitors is shown in Figure~\ref{fig:let:all_gains}.

\begin{figure}[htb] 
   \centering
   \includegraphics[width=2.9in]{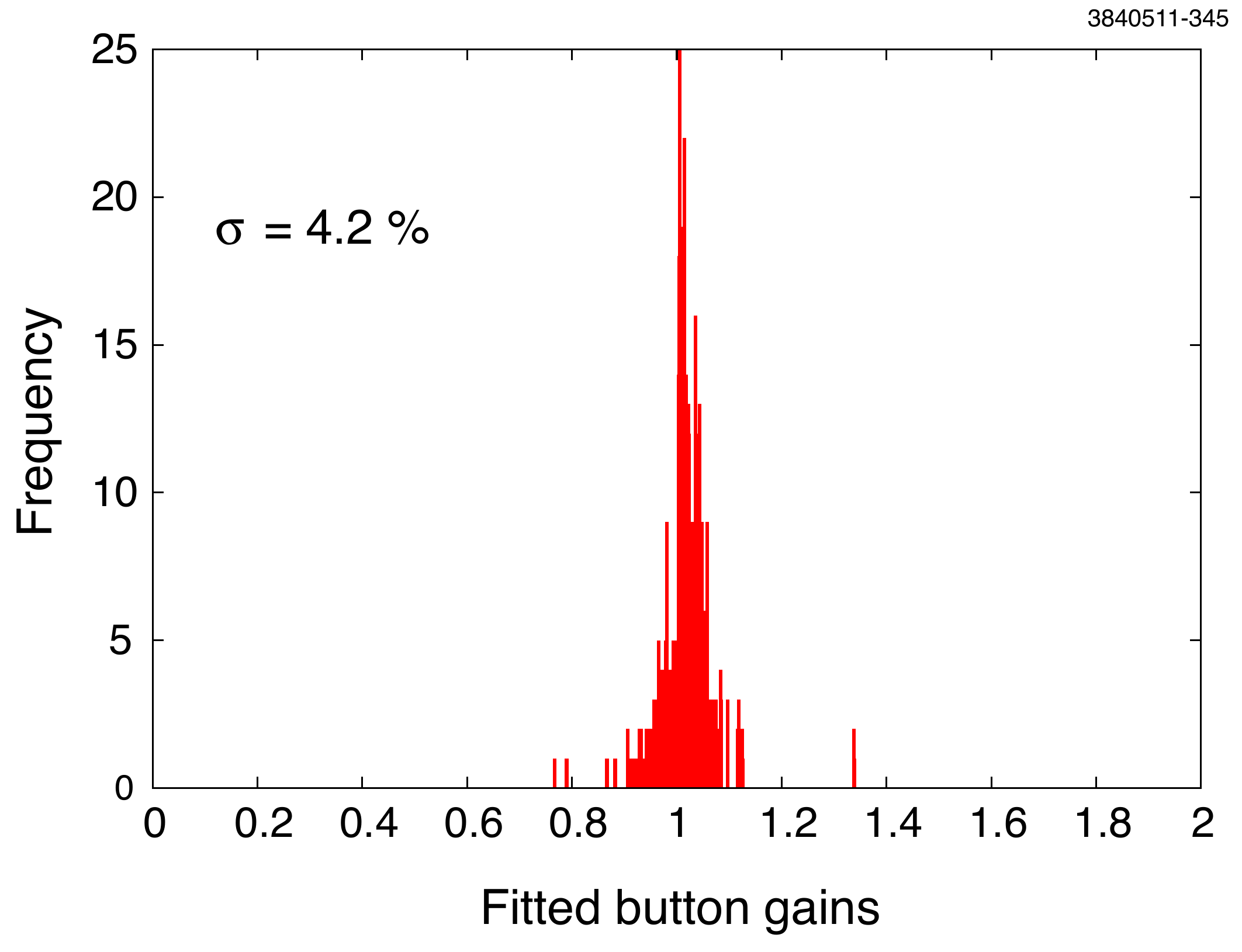}
   \includegraphics[width=2.9in]{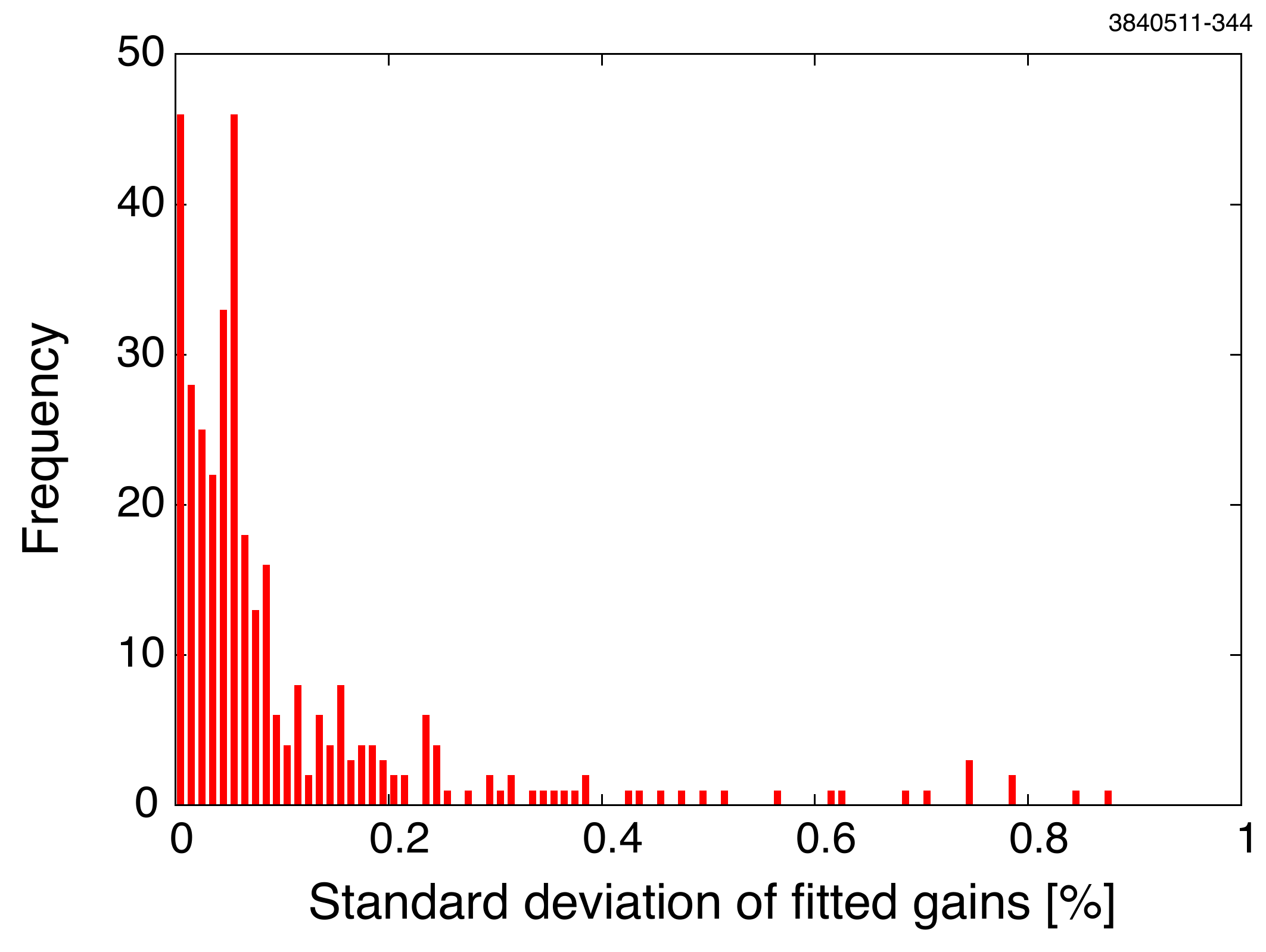}
  \caption[Distribution of measured
    button gains.]{\label{fig:let:histogram_gains} (Left) Distribution of measured
    button gains, with standard deviation 4.2\%. (Right) Distribution of the
    variation in fitted
    button gains for seven sets of turn-by-turn data.}
\end{figure}

A histogram of the individual gains is shown in left-hand plot of
Figure~\ref{fig:let:histogram_gains}. The accuracy of the beam based
method for determining button gains is reflected in the
reproducibility of the fitting procedure.  The right-hand plot of
Figure~\ref{fig:let:histogram_gains} shows the distribution of the
variation in fitted gains for seven distinct sets of turn-by-turn
data. From this one can conclude that the gains are characterized to
within a fraction of a percent precision.

\subsection{BPM Tilts}
As described above BPM tilts couple measured horizontal and vertical
dispersion. In particular for a BPM horizontal-to-vertical tilt
$\theta$, $\Delta\eta_y \sim \eta_x\theta$. Peak horizontal
dispersion in the CesrTA lattice is about 1-2~m. In order that the
error in the measured vertical dispersion be less than 10~mm, we
require that uncertainty in the BPM tilts be less than 10~mrad. BPM
tilts can be extracted from the coupling measurements by taking
advantage of the fact that the out of phase component of the
coupling matrix $\overline C_{12}$ is insensitive to tilt. Then if
$\overline C_{12}$ is corrected with skew quads, the in phase
components $\overline C_{11}$ and $\overline C_{22}$ provide a
direct measure of the physical tilt.

\begin{figure}[htb] 
   \centering
   \includegraphics[width=2.9in]{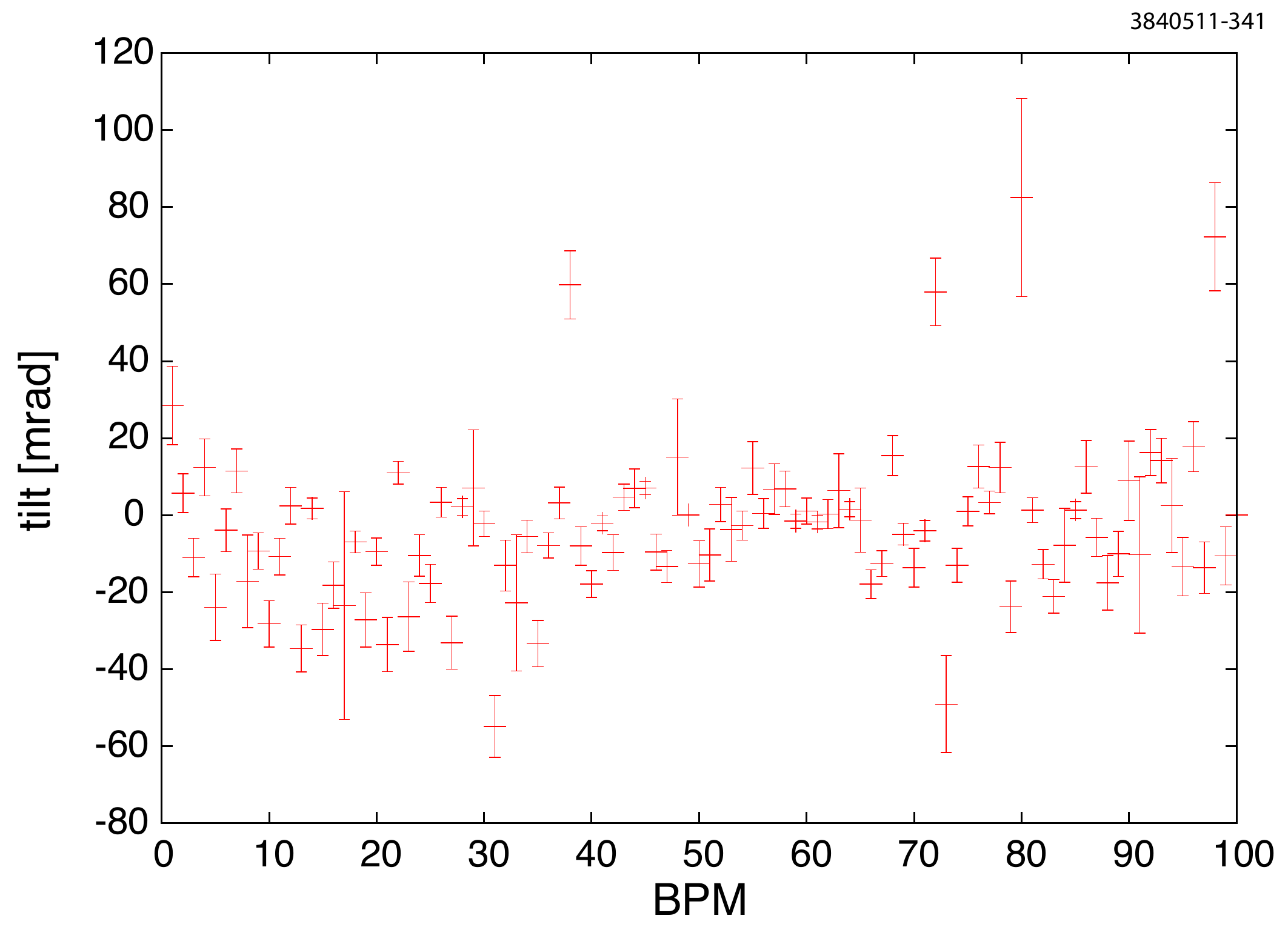}
   \includegraphics[width=2.9in]{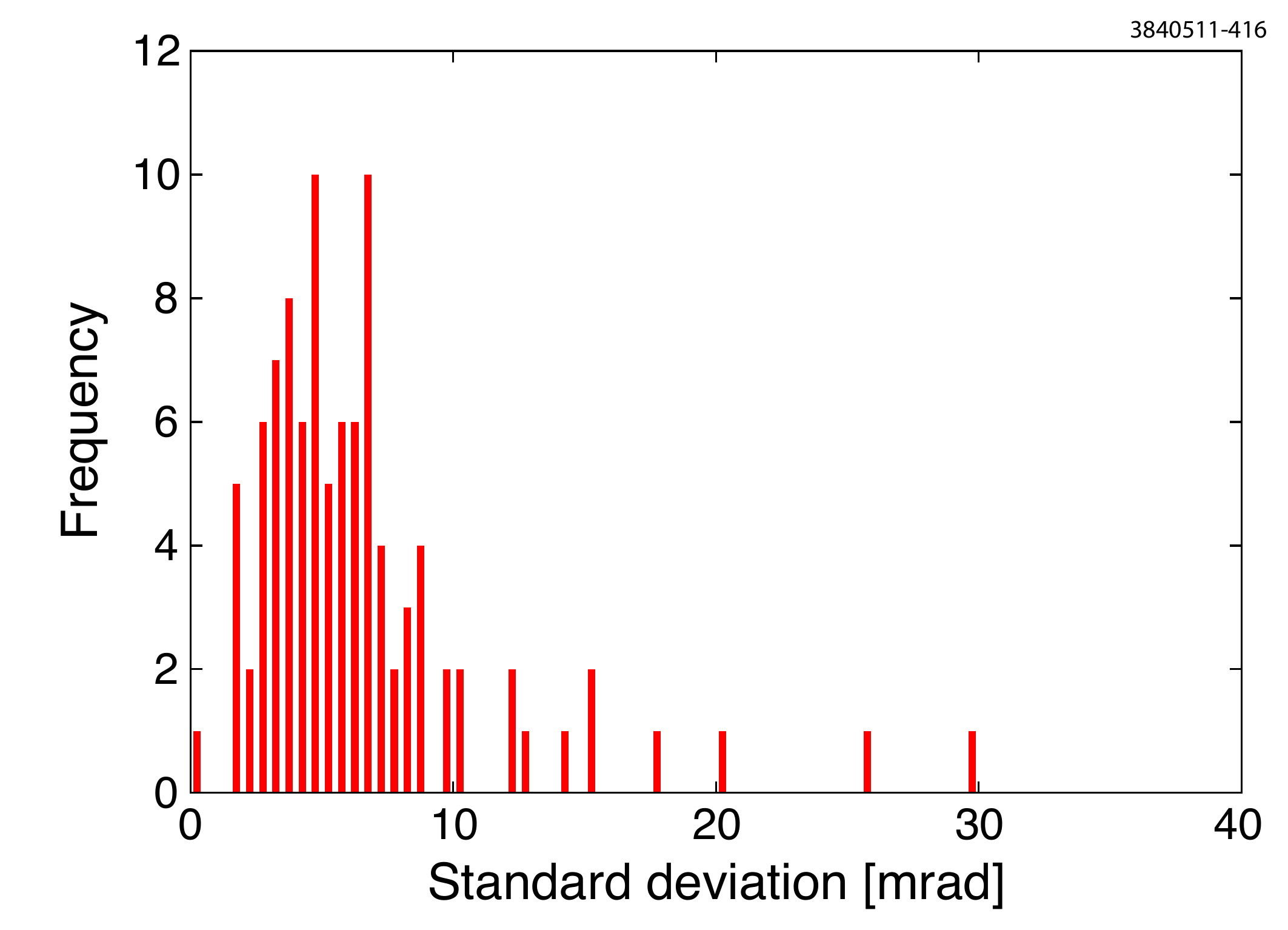}
  \caption[Measured BPM
    tilts.]{\label{fig:let:all_tilt} (Left) Measured BPM
    tilts. The error bars are $\sigma={1\over N}\sqrt{\sum_{i=1}^N (\theta
      - \langle\theta\rangle)^2}$ of 58 data
    sets. (Right) The distribution of the RMS residuals.}
\end{figure}

The algorithm for extracting BPM tilt from the coupling data is as
follows:
\begin{enumerate}
\item Measure and correct coupling based on $\overline  C_{12}$
\item Remeasure coupling, specifically $\overline  C_{11}$ and $\overline
  C_{22}$.
\item Model the measured  $\overline  C_{11}$ and $\overline C_{22}$ by fitting
  BPM tilts.
\end{enumerate}
The procedure has been applied to 58 sets of coupling data, all
collected during the course of the June 2011 run. The 58 data sets
include measurements in five distinct lattice configurations and at
four different beam energies. The average along with the rms of the
residuals of the fitted tilts are shown in left-hand plot of
Figure~\ref{fig:let:all_tilt}. A histogram of the RMS of the
residuals is shown in right-hand plot of
Figure~\ref{fig:let:all_tilt}. Uncertainty in the measured tilts is
less than 10 mrad.

\section[Performance of Low Emittance Instrumentation]{Performance of Low Emittance Instrumentation}
\label{sec:let.let_instrumentation}

\subsection{Orbit}
\begin{figure}[htb] 
   \centering
   \includegraphics[width=2.9in]{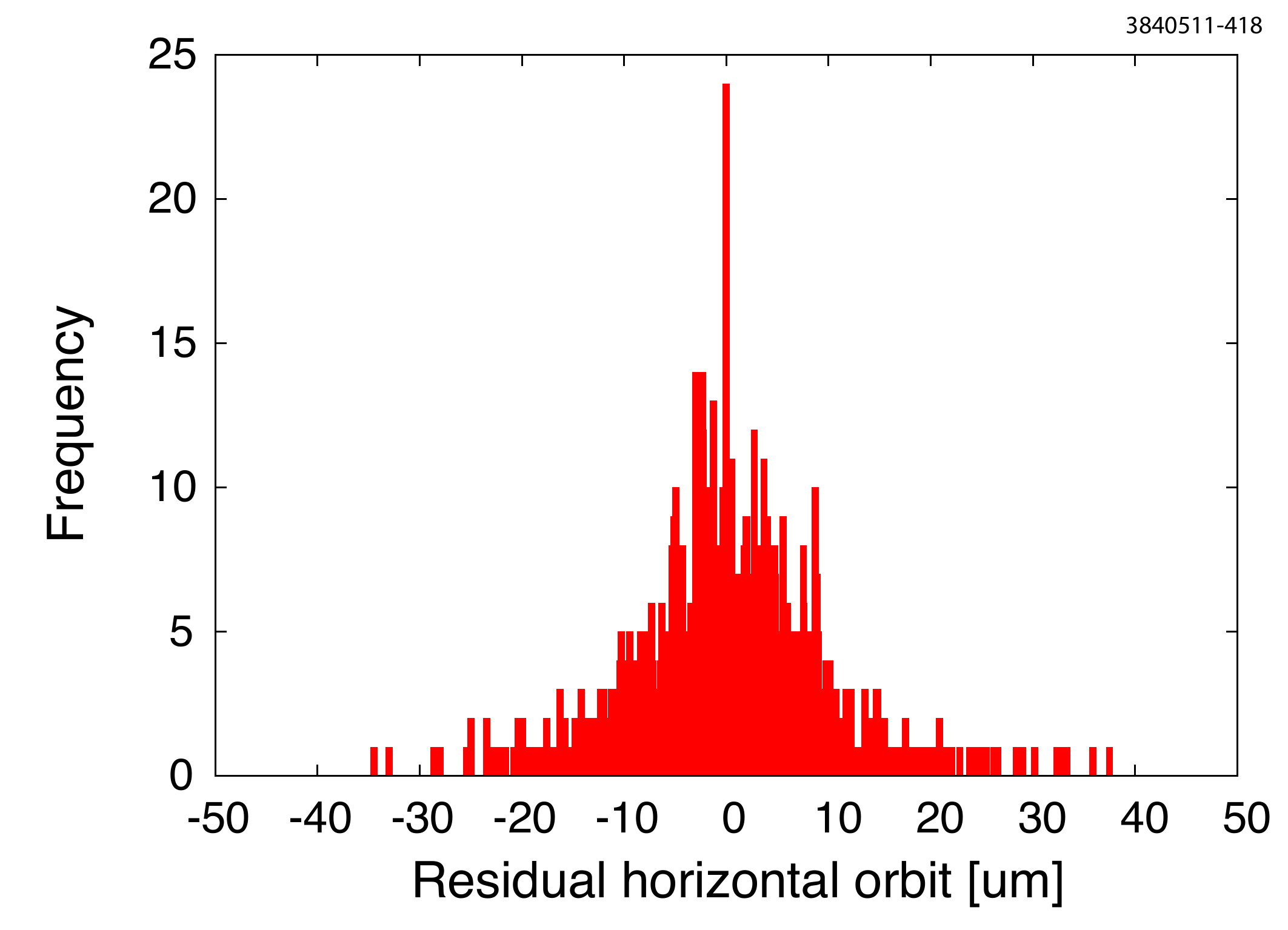}
   \includegraphics[width=2.9in]{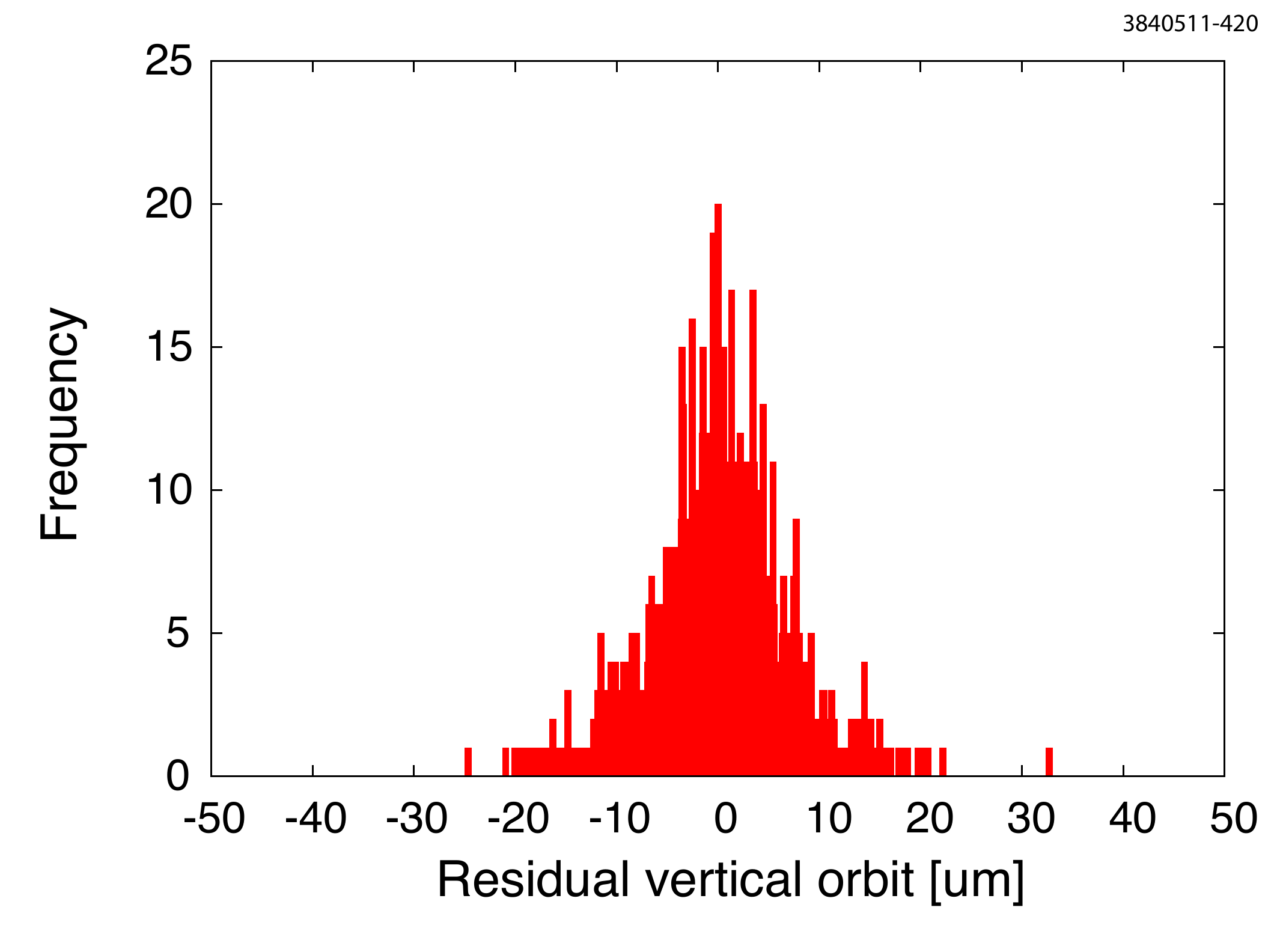}
  \caption[Reproducibility of
    horizontal and vertical orbit measurements.]{\label{fig:let:orbit_reproducibility} Reproducibility of
    horizontal (left) and vertical (right) orbit measurements. The
    standard deviation of the measurements is $7.8\,\mu$m and $5.8\,\mu$m, respectively.}
\end{figure}

We estimate the intrinsic resolution of the orbit measurement in
terms of reproducibility. The variation in the measured positions
for twenty consecutive measurements is shown in
Figure~\ref{fig:let:orbit_reproducibility}. The standard deviation
of horizontal and vertical position measurements is $7.8\,\mu$m and
$5.8\,\mu$m, respectively. This does not follow the expected
$1/\sqrt{N}$ behavior for random noise, which implies true beam
motion is contributing to the orbit measurement resolution. The
source of this beam motion is not diagnosed at the time of this
publication, and efforts continue on this issue.

\subsection{Dispersion}
\begin{figure}[htb] 
   \centering
   \includegraphics[width=2.9in]{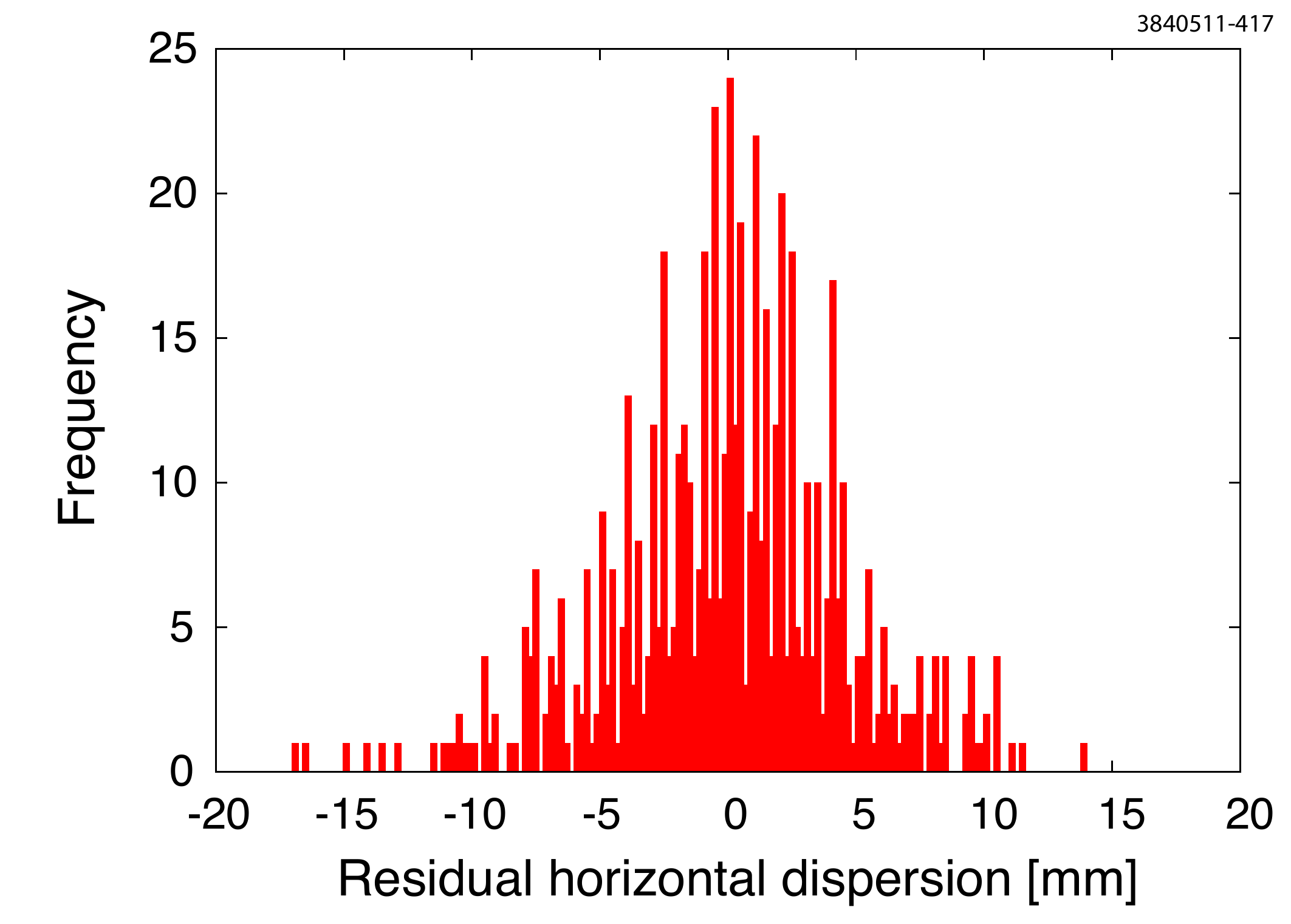}
   \includegraphics[width=2.9in]{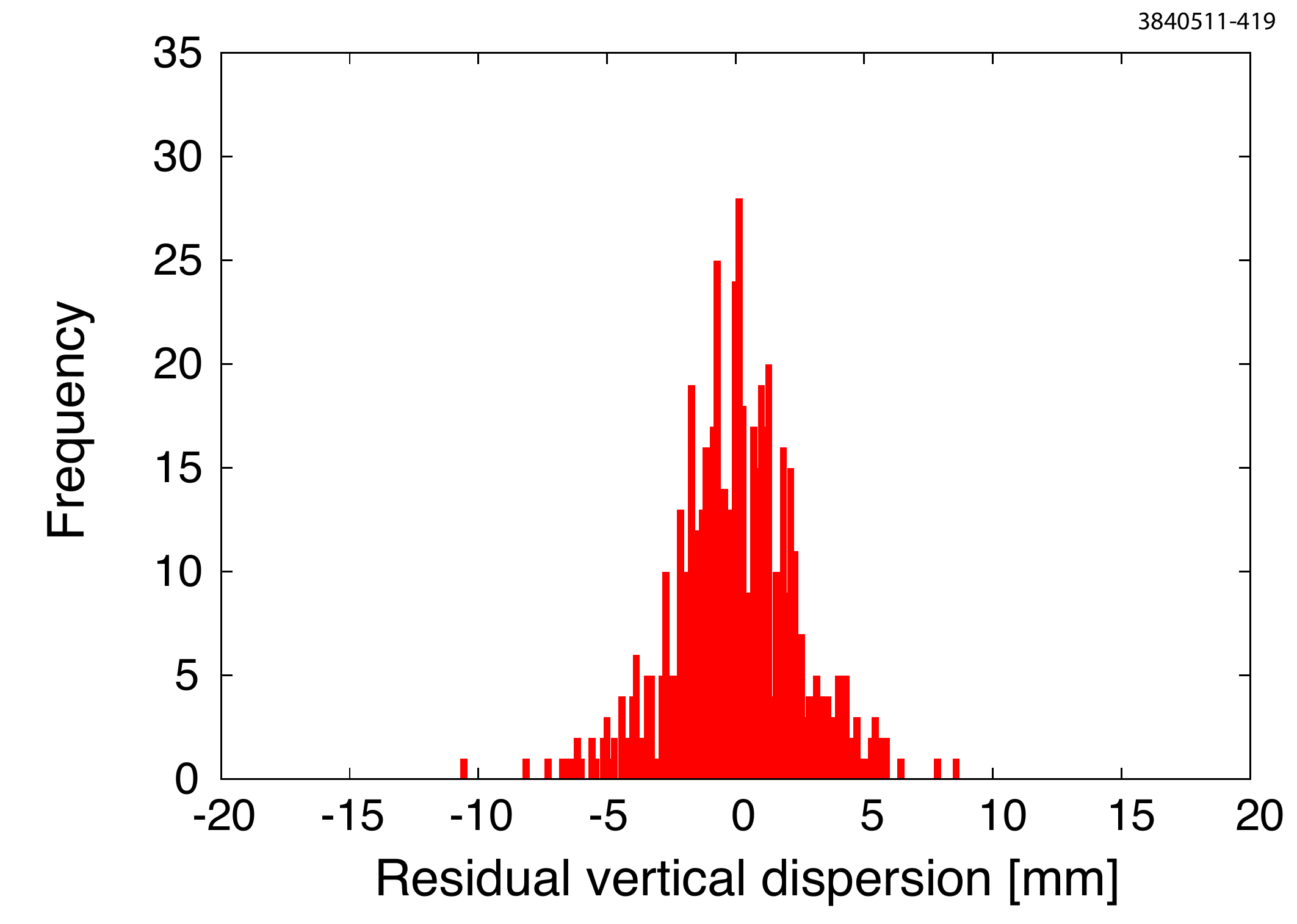}
  \caption[ Reproducibility of
    horizontal and vertical dispersion measurements.]{\label{fig:let:eta_reproducibility}
    Reproducibility of
    horizontal (left) and vertical (right) dispersion measurements. The
    standard deviation of the measurements is $4.6$ mm and $2.3$ mm respectively.}
\end{figure}

The distribution of the variation of the measured horizontal and
vertical  dispersion is shown in
Figure~\ref{fig:let:eta_reproducibility}.  The standard deviations
are 4.6 mm for horizontal and 2.3 mm for vertical dispersion,
corresponding to the intrinsic resolution of the dispersion
measurement. The accuracy of the dispersion measurement is limited
by the systematics described above, rather than the intrinsic
resolution of the beam position monitors.
\subsection{Phase and Coupling}
\begin{figure}[htb] 
   \centering
   \includegraphics[width=2.9in]{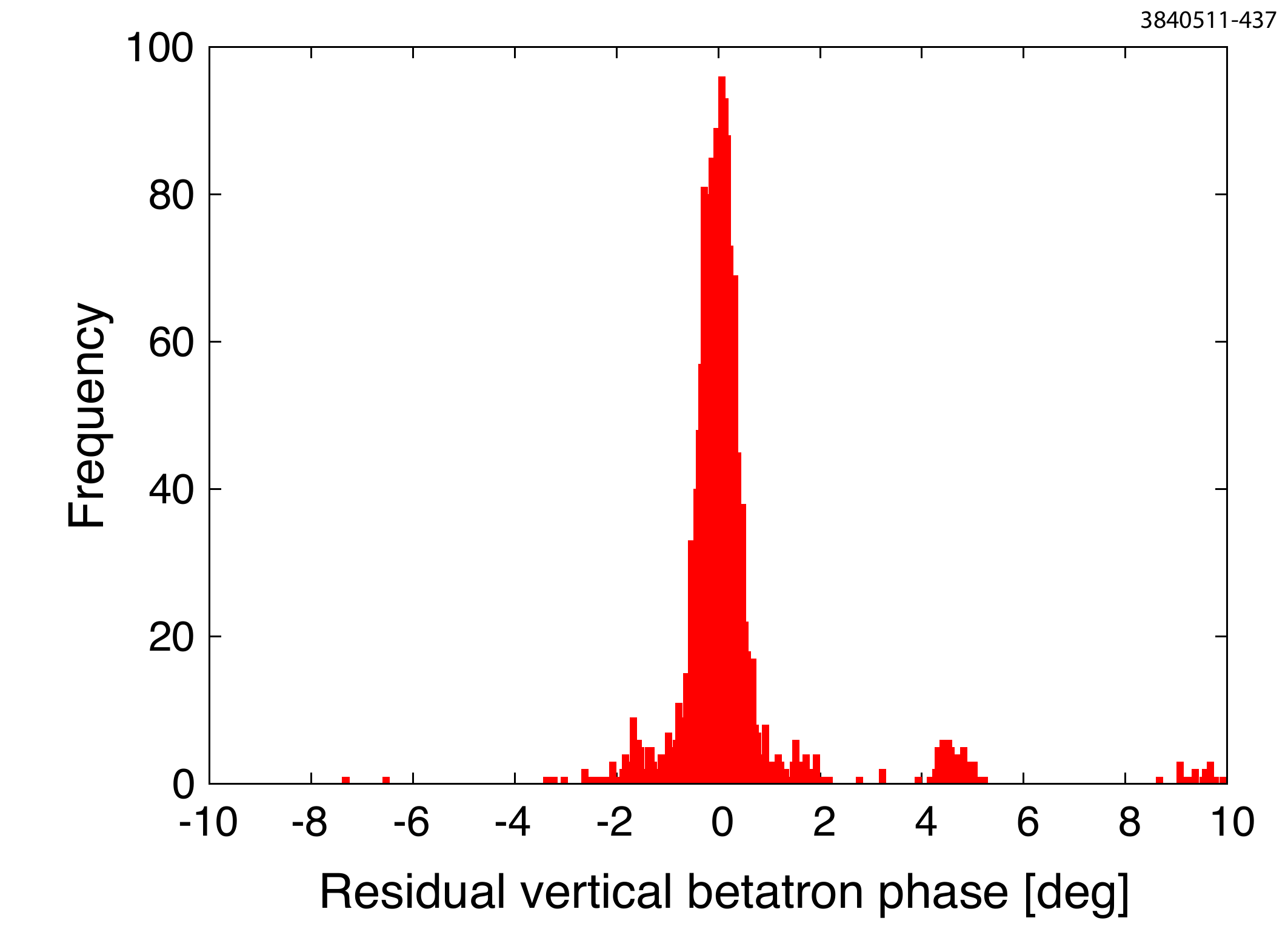}
   \includegraphics[width=2.9in]{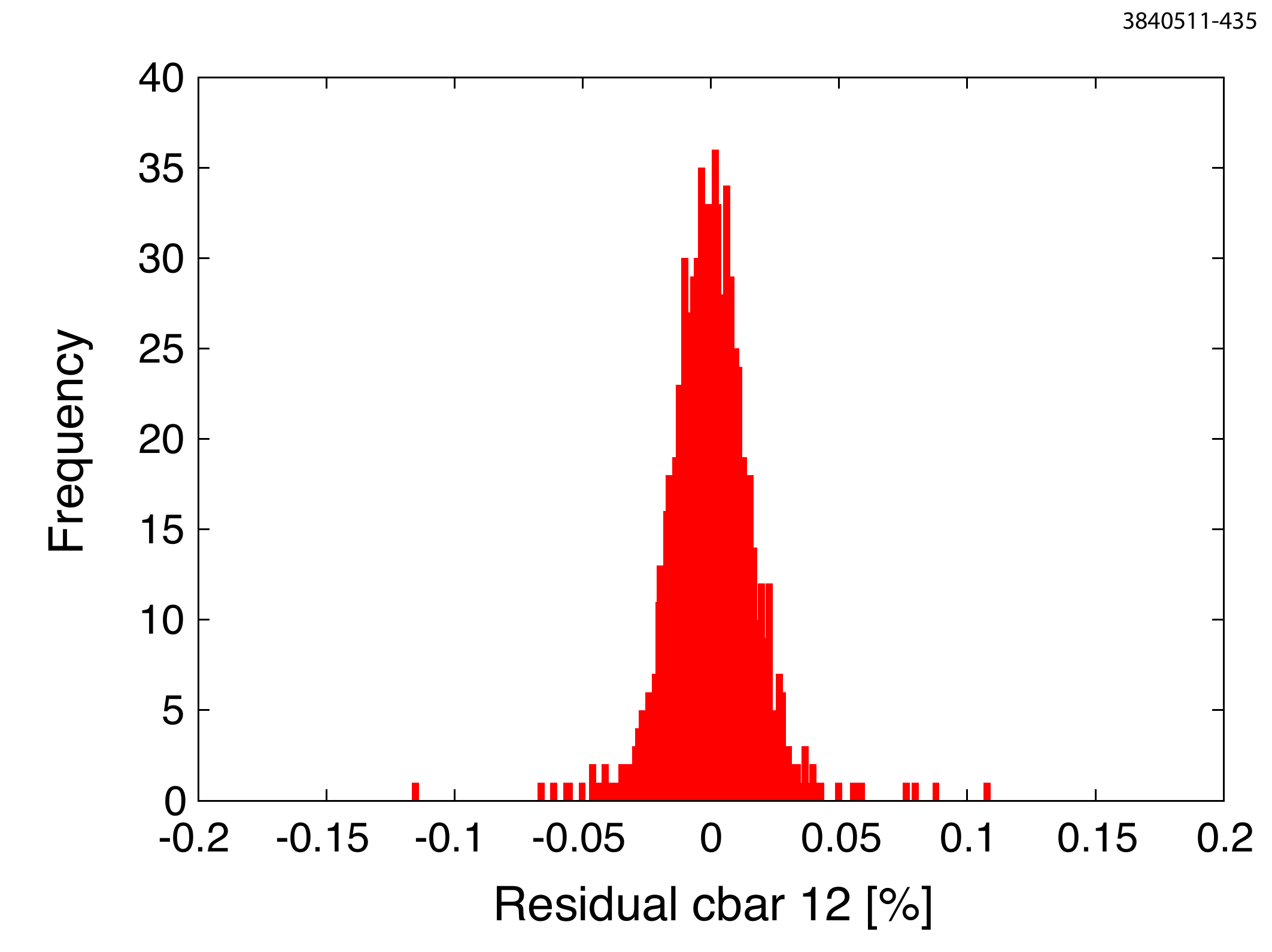}
  \caption[Reproducibility of vertical betatron phase
    measurements and transverse coupling.]{\label{fig:let:phase_reproducibility} Reproducibility of
     vertical betatron phase
    measurements (left) and transverse coupling element $\overline C_{12}$ (right).}
\end{figure}

In order to determine the intrinsic resolution limit of the phase and
coupling measurement, we repeat the measurement 20 times
consecutively. Differences from one measurement to the next are due
either to drift in machine parameters or statistical limitations of the
technique. The distribution of the residuals with respect to the
average measurement of vertical phase for each BPM is shown in the top plot of
Figure~\ref{fig:let:phase_reproducibility}, and the distribution of residuals
with respect to the average coupling ($\overline C_{12}$) are shown in the bottom plot of
Figure~\ref{fig:let:phase_reproducibility}.


\subsection{xBSM - Vertical Beam Size}
\label{ssec:let.let_instrumentation.xbsm} The xBSM images
synchrotron radiation x-rays emitted in a hard bend dipole on a
bunch-by-bunch, turn-by-turn basis using a linear array of 32
photodiodes \cite{NIMA798:127to134}. The xBSM is the principle tool
for measuring vertical emittance and determining the effectiveness
of the low emittance tuning procedures.

Several optics are available for the xBSM. A simple pinhole image is
the most straightforward to interpret and is the basis for most of
the low emittance tuning studies. (The optic that is described as
the ``pinhole'' is in reality a vertical gap with adjustable
height.) Neglecting diffraction, light from a point source passing
through a slit of height $G$ produces an image of width
$\sigma_{point source} = \frac{G}{x_s}(x_i+x_s)$, where $x_i$ and
$x_s$ are distances from optic-to-image and source-to-optic,
respectively. The ``effective height'' of the pinhole can then be
taken to be $\sigma_p = \sigma_{point source} / M$, where $M =
x_{i}/x_{o}$ is the magnification of the pinhole. The pinhole (gap
height) is set to give minimum image size for the mean x-ray energy
of 2 keV (at 2.085 GeV beam energy). The effective height, $\sigma_p
= 21 \pm 2\,\mu$m, is defined as the convolution of magnified gap
image and diffraction distribution.

The final beam image will be computed as the convolution of the
image of the source from an infinitely small pinhole and the image
of a finite-height gap from a point source. Details of how the xBSM
data is post-processed into a beam size are available in
\cite{NIMA798:127to134}. Systematic and statistical error analysis
is shown in \cite{CornellU2013:PHD:JShanks, PRSTAB17:044003}.

\subsection{Summary of Low Emittance Measurement Resolution}

The resolution of measurements of beam position, dispersion,
coupling, betatron phase advance, and vertical beam size are derived
in \cite{CornellU2013:PHD:JShanks, PRSTAB17:044003}, and are
summarized in Table \ref{tab:let:measurement-resolution}.

\begin{table}[htb]
\begin{center}
\caption[Accuracy of beam based
measurements]{\label{tab:let:measurement-resolution}Accuracy of beam
based measurements. Vertical emittance resolution is detailed in
\cite{PRSTAB17:044003}} \vspace*{1ex}
\begin{tabular}{|c|c|}
\hline
$\;$ Parameter & $\;$ RMS $\;$ \\
\hline
  BPM position (absolute)  [$\mu$m]                   &  170  \\
  BPM position (differential) [$\mu$m]                &  10   \\
  Dispersion [mm]                                     &  10   \\
  Coupling($\overline C_{12}$) [\%]                    &   0.2 \\
  Betatron phase [deg]                                &   0.1 \\
  Vertical beam size [$\mu$m]                         &   2.5 \\
  Vertical emittance (statistical / systematic) [pm]  &  $\pm$ 3.5 /
  $\pm$ 0.3 \\
\hline
\end{tabular}
\end{center}
\end{table}



\section[Electron Cloud and Intra-beam Scattering Experimental Hardware and Techniques]{Electron Cloud and Intra-beam Scattering Experimental Hardware and Techniques}
\label{sec:ec_dynamics.hardware_techniques}

This section describes the measurement techniques that are required
for EC dynamics and intra-beam scattering measurements in {CesrTA}.
The requirements for the single-bunch intra-beam scattering are
generally a subset of those employed for the multi-bunch EC dynamics
measurements. (One notable exception to this is that intra-beam
scattering measurements include streak camera measurements for
determining bunch lengthening and these are not routinely employed
for EC dynamics measurements.) As a result this section will focus
on the techniques for EC dynamics measurements exclusively with the
understanding that many of these will also be utilized for
intra-beam scattering measurements.

\subsection[Requirements for Electron Cloud Studies]{Requirements for Electron Cloud Induced Beam Studies}
\label{ch:ec_dynamics}

One of the key goals of the {CesrTA} research program has been to
improve the understanding of the interaction of the EC with the high
energy particle beam.  The initial motivation for this was to allow
the extrapolation from the experimental conditions of {CesrTA} to
the conditions anticipated for the ILC damping rings.

The interaction of the particle beam with the cloud can be studied
by measuring the properties of the beam in the presence of the
cloud. The key beam properties which are influenced by the cloud are
the beam's closed orbit distortion (quite small, and not extensively
studied with {CesrTA}), the frequency spectrum of the beam
centroid's coherent dipole motion relative to this orbit and the
beam's transverse position distribution.

In {CesrTA}, the beam is configured longitudinally into a train of
$\sim 10$ mm long bunches separated by an adjustable spacing
(variable from a minimum of 4~ns, up to a maximum equal to the
revolution period, 2.56~$\mu$s). For sufficiently closely-spaced
bunches, the EC grows along the train and so the cloud environment
is different for each bunch. For this reason, it is critical that
the beam dynamics measurement made to probe the cloud be done on a
bunch-by-bunch basis.

The  frequency spectrum of the coherent motion of each bunch contains a wealth of information. In particular,  this information includes
\begin{itemize}
\item the amplitude, frequency, and line shape of the betatron dipole lines, which are sensitive to the electron cloud's electric field,
to the mode of oscillation of the bunches in the train, and to the presence of multi-bunch instabilities;
\item  the amplitude, frequency, and line shape of head-tail lines, which are generally separated from the betatron lines by
approximately the synchrotron frequency, and are sensitive to internal motion within the bunch driven by EC-induced single-bunch head-tail instabilities.
\end{itemize}
In addition after bunch motion has been excited by an external
source, the time dependence of the amplitudes of the betatron dipole
and head-tail lines provide information on the damping of these
lines, which is related to aspects of the effective EC impedance
that is not probed by tune measurements. In the sections below the
experimental hardware and techniques used to obtain the measurements
are described and the some typical beam dynamics observations are
presented.


There are several beam parameters particularly relevant for the
study of electron cloud effects. Since the EC can produce focusing
of the stored beam, measuring the betatron tunes of bunches along
the train gives information about the average density of the cloud
along the length of the train.  The electron cloud can also produce
unstable motion in later bunches in the train.  To observe the
unstable motion, it is necessary to detect the amplitude at the
betatron frequency and any other frequencies representing different
modes of oscillation (e.g. head-tail modes) of bunches within the
train.  The unstable motion may also result in enlargement of the
vertical beam size, thus the measurement of the vertical beam size
for each bunch in the train is important.  Before beam conditions
approach the regime for the onset of unstable motion, it is possible
to measure the damping of coherent motion of the bunches using
drive-damp techniques.  This method excites coherent dipole betatron
modes or head-tail modes for each bunch within the train and then
observes the damping of the motion.  Thus it is possible to observe
how the coherent motion becomes less stable before the onset of
instability. The methodology and examples of typical measurements
for these techniques are presented in the following sections.

\subsection{Bunch-by-Bunch Tune Measurements}
\label{ssec:ec_dynamics.hardware_techniques.tunes}

During the course of the {CesrTA} project several different
techniques have been utilized for making tune shift measurements for
individual bunches within trains of bunches.  These techniques,
their benefits and their limitations will be described in this
section.

\subsubsection{Multi-bunch Large Amplitude Excitation}
\label{sssec:ec_dynamics.hardware_techniques.tunes.mblae}

This method for observing the tunes of different bunches within the
train consists of pulsing a pinger magnet with a single-turn
excitation to deflect all of the bunches within the train, thereby
starting a dipole oscillation of their centroids.  The CBPM system
is then timed to read out a number of BPMs over several thousand
turns for all bunches in the train; the data acquisition is
synchronized with the triggering of the pinger magnet's deflection.
Turn-by-turn trajectory data are recorded, typically for 8192 turns.
Note that the damping time depends on a bunch's position within the
train; bunches at the start of the train have transverse damping
times of 10-20,000 turns, whereas bunches later in the train have
shorter damping times due to their interaction with the electron
cloud. The data are analyzed offline with a Fast Fourier transform
(FFT) algorithm, from which the betatron tunes are determined. For
8192 turns, the FFT has frequency resolution $f_{rev}/8192 =
390.1~\textrm{kHz}/8192 = 48~\textrm{Hz}$; more advanced methods
have improved resolution. During many of these measurements the peak
horizontal and vertical beam displacements were typically 5~mm at
2.1~GeV and 2~mm at 5.3~GeV.

Since data from all bunches is recorded at the same time, it is
relatively rapid to take data in one set of conditions and, since
the data from all bunches is taken on the same turns, this method is
relatively insensitive to any drifts in the storage ring tunes.
However, the fact that all bunches are excited at the same instant
implies that the lowest coupled-bunch mode is necessarily excited
for the train of bunches.  When the train is oscillating in this
mode, the bunch-by-bunch horizontal tune shifts induced by the EC
are strongly suppressed, and difficult to measure.  It is also the
case that the pinger excitations are relatively large with respect
to the stored beam's size: e.g., typically the vertical oscillation
amplitude may exceed several ten's of vertical sigma.  So the beam's
oscillation is exploring a fairly large volume of the electron cloud
charge distribution.

\subsubsection{Single Bunch Small Amplitude Excitation}
\label{sssec:ec_dynamics.hardware_techniques.tunes.sbsae} 
\begin{figure}[htbp] 
   \centering
   \includegraphics[width=0.6\columnwidth]{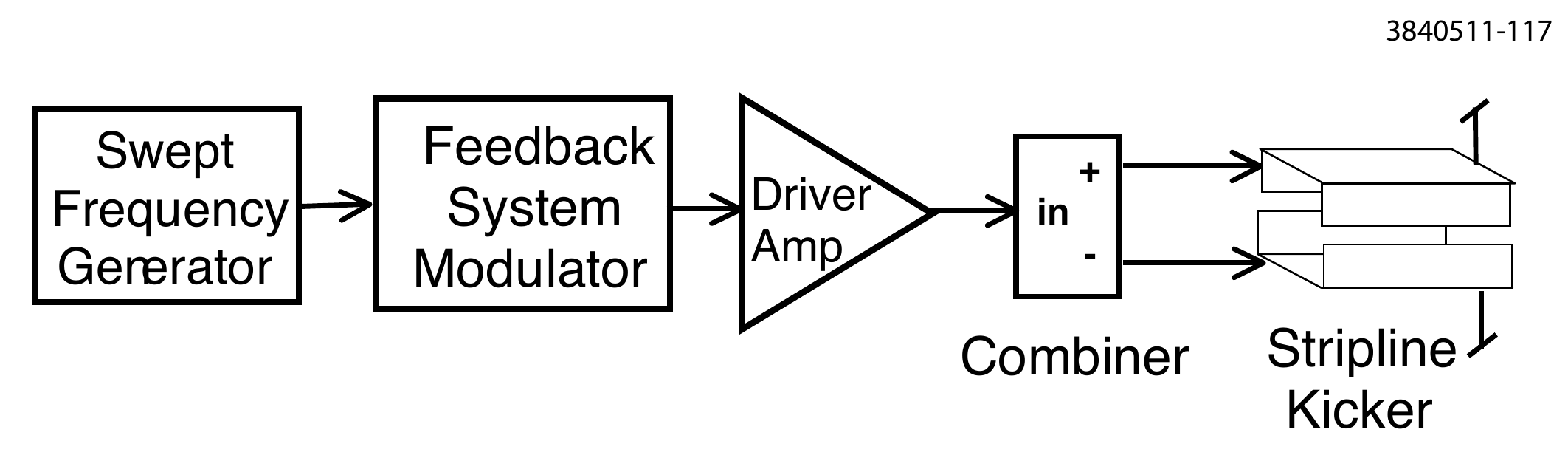}
   \caption[Stripline driver for single bunch excitation]{\label{fig:SingleBunchDriverDiagram}
   Single bunch excitation method using the stripline kicker, driven by a swept frequency source via the feedback system's external modulation port. }
\end{figure}

Another
approach has been developed for bunch-by-bunch tune measurements.
This approach excites only a single bunch in the train, thereby
reducing the coupling from earlier bunches to the bunch that is
being measured.  This is accomplished by driving both the horizontal
and vertical stripline kickers (shown schematically for one
stripline kicker in Fig.~\ref{fig:SingleBunchDriverDiagram}), using
the external modulation input for the beam stabilization feedback
system, which allows gating of the input signal into the appropriate
timing window in order to excite only the bunch being measured.  The
source for the signal for the external modulation port comes from a
frequency synthesizer, whose output frequency is swept across the
range of betatron oscillation frequencies, covering the tunes for
the entire ensemble of bunches.  The frequency is swept with a
saw-tooth at 500~Hz, driving the bunch in its dipole oscillation
mode when the excitation frequency crosses the betatron resonance.
The turn-by-turn position data are recorded for a number of BPMs
using the CBPM system readout with the total number of turns
sufficiently large to capture at least one full excitation and
damping cycle. The measurement process is repeated as the
excitation's delay is stepped from one bunch to the next, resulting
in a set of positions for all bunches at each delay.  The data are
analyzed offline with an FFT to give the oscillation frequency of
the excited bunch and coupling of its motion to subsequent bunches
via the EC.

\begin{figure}[htbp] 
   \centering
   \includegraphics[width=3in]{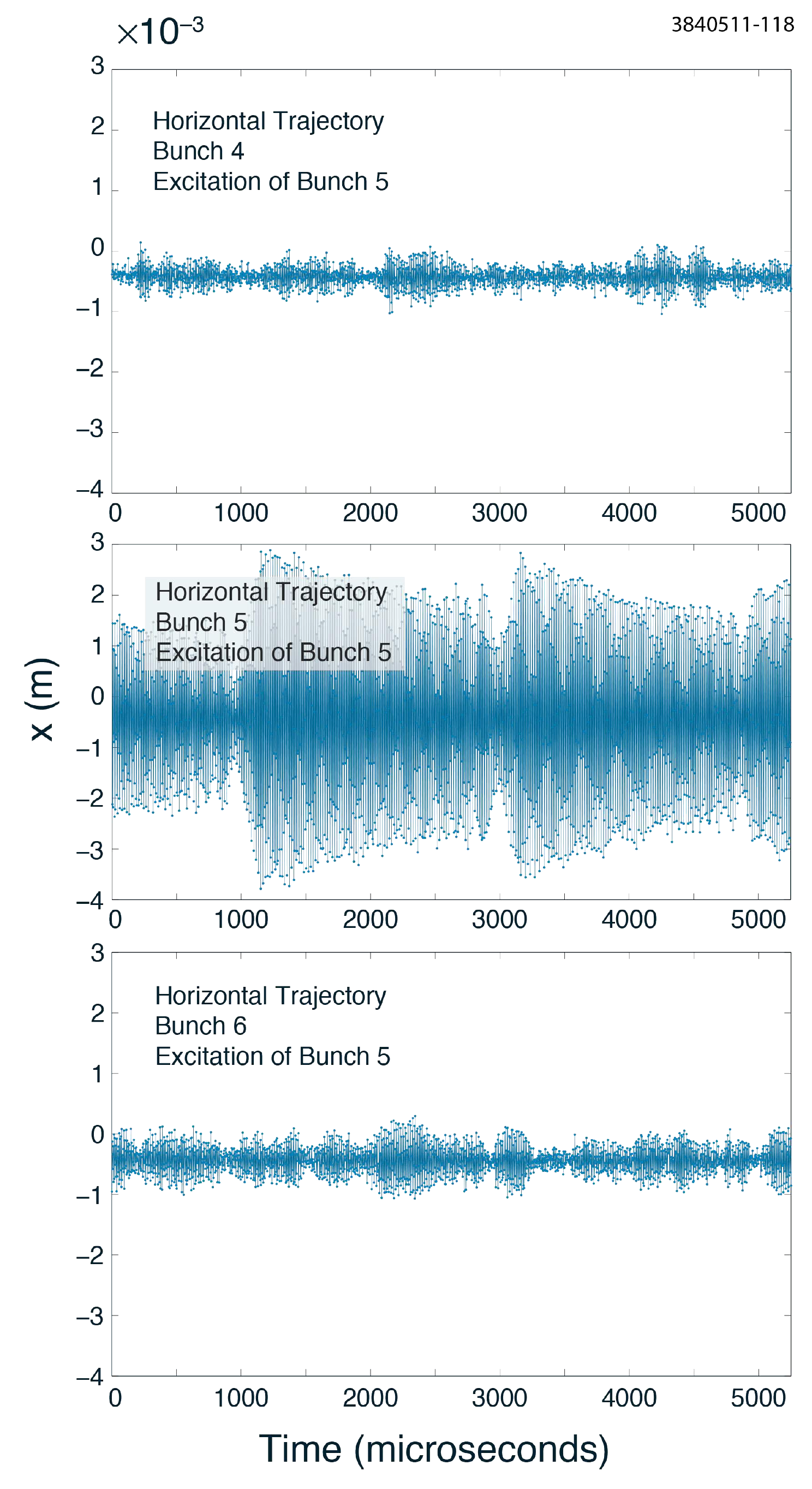}
   \caption[EC single bunch horizontal position data]{\label{fig:SingleBunchHorzPositionData}
   Horizontal position of bunches 4, 5 and 6 (respectively for the top, middle and bottom plots) for a 10-bunch train when only bunch number 5 was excited.}
\end{figure}

\begin{figure}[htbp] 
   \centering
   \includegraphics[width=3in]{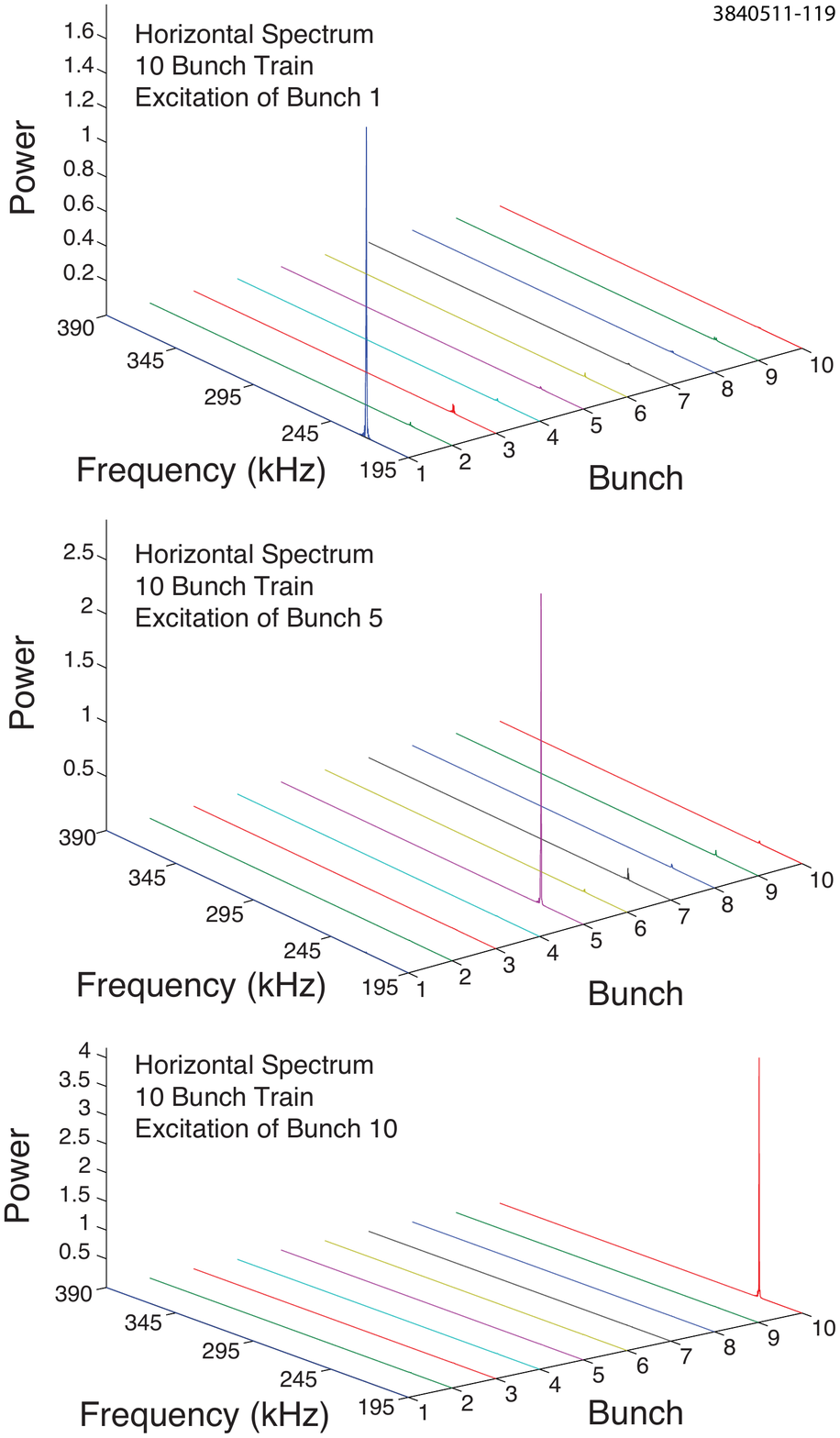}
   \caption[EC single bunch horizontal position FFT]{\label{fig:SingleBunchHorzPositionFFT}
  Horizontal position spectra of all bunches in a 10-bunch train when bunches number 1, 5 and 10 (respectively for the top, middle and bottom plots) were driven individually. }
\end{figure}

Some results are presented here to illustrate this technique; the
data were taken with a 10~bunch train having a 14~ns spacing in
2.1~GeV beam energy conditions.
Fig.~\ref{fig:SingleBunchHorzPositionData} shows the horizontal
position data for the fourth, fifth and  sixth bunch, when only
bunch number 5 was being excited.  During the 2048-turns of the
data-samples taken on simultaneous turns for the three bunches, it
is clear that bunch~5 was excited with two complete cycles of the
swept signal source.  This is even clearer in
Fig.~\ref{fig:SingleBunchHorzPositionFFT} which shows the horizontal
spectra of all 10~bunches when bunch numbers 1, 5 and 10 were being
driven individually.  The fact that the stripline kicker is driving
only one bunch is quite evident in both
Figures~\ref{fig:SingleBunchHorzPositionData} and
~\ref{fig:SingleBunchHorzPositionFFT}.

\begin{figure}[htbp] 
   \centering
   \includegraphics[width=3in]{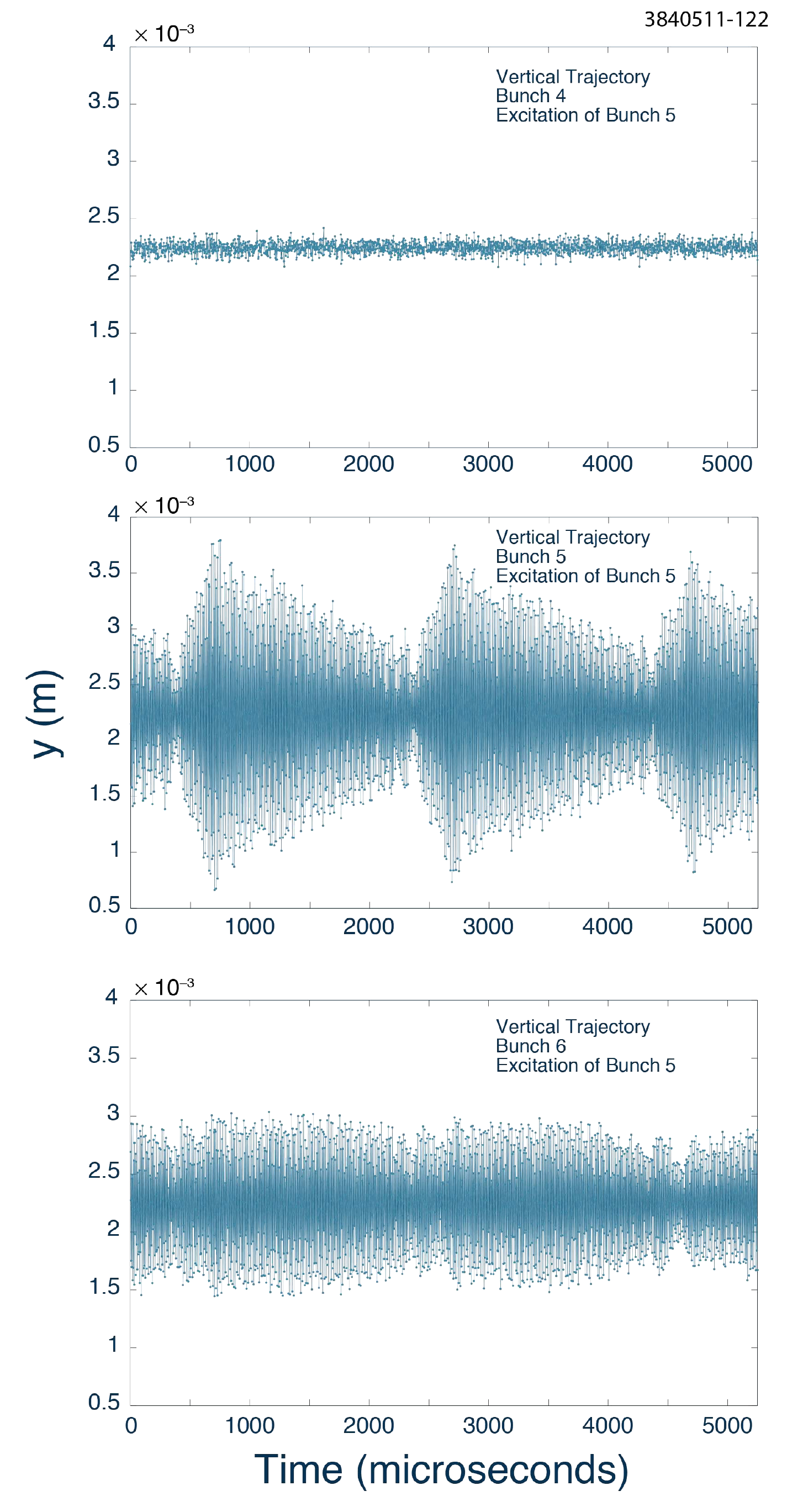}
   \caption[EC single bunch vertical position data]{\label{fig:SingleBunchVertPositionData}
   Vertical position of bunches 4, 5 and 6 (respectively for the top, middle and bottom plots) for a 10-bunch train when only bunch number 5
   was excited. Turn-by-turn position indicated by red dots, connected by blue lines.}
\end{figure}

\begin{figure}[htbp] 
   \centering
   \includegraphics[width=3in]{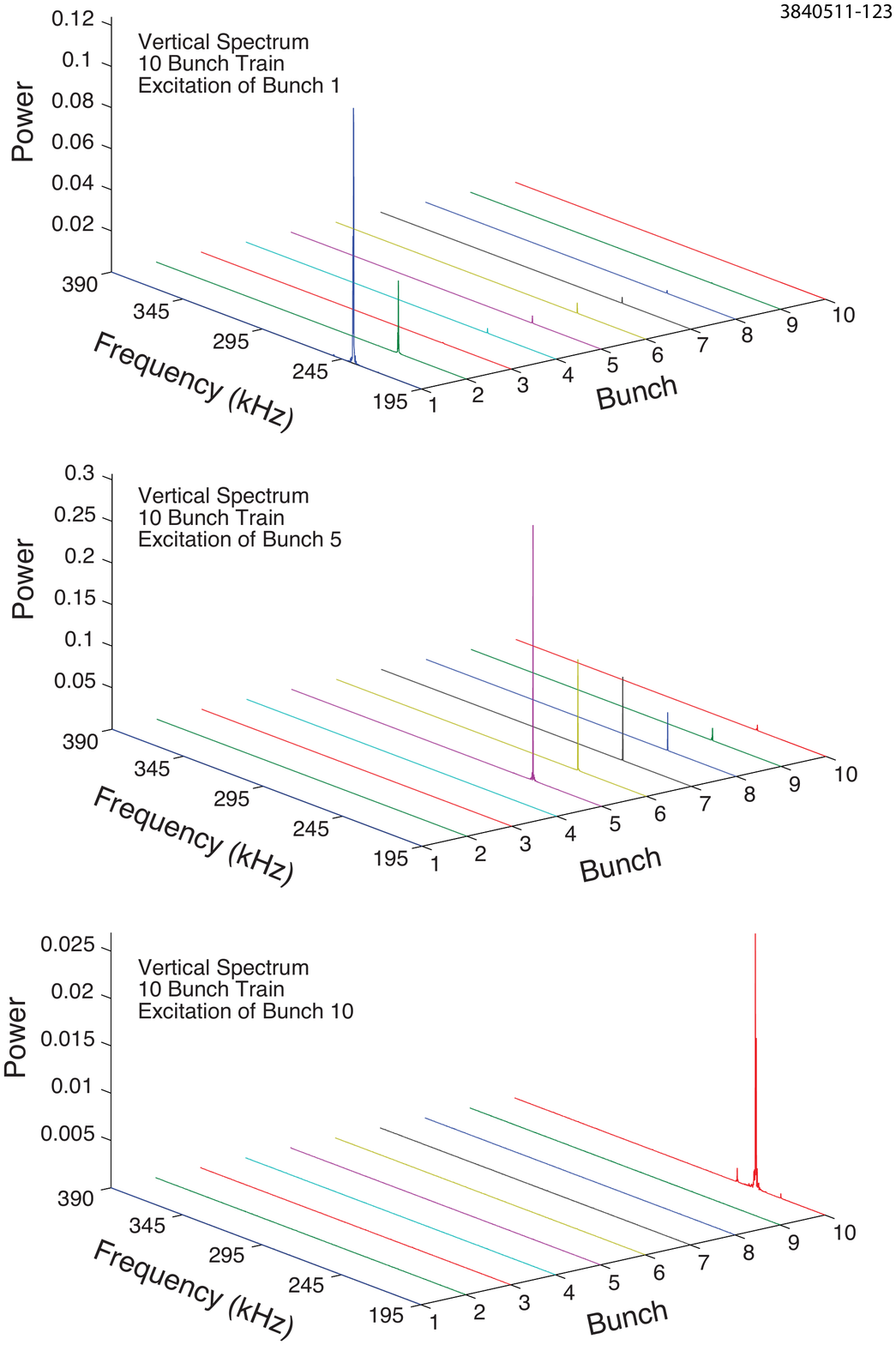}
   \caption[EC single bunch vertical position FFT]{\label{fig:SingleBunchVertPositionFFT}
   Vertical position spectra of all bunches in a 10-bunch train when bunches number 1, 5 and 10 (respectively for the top, middle and bottom plots) were driven individually. }
\end{figure}

For comparison with the horizontal data, the matching set of
vertical data are presented here for the same storage ring and EC
conditions as above.  The vertical position data for bunches 4, 5
and 6 is shown in Fig.~\ref{fig:SingleBunchVertPositionData}, when
only bunch~5 is driven.  Also the vertical spectra for all bunches
are shown in Fig.~\ref{fig:SingleBunchVertPositionFFT}, when bunches
1, 5 and 10 are individually excited.  An interesting feature,
visible in the vertical data, is that even though only one bunch is
being driven, its motion couples to subsequent bunches in the train.
Fig.~\ref{fig:SingleBunchVertPositionFFT} presents evidence that
this coupling increases along the train, suggesting that the EC may
be playing a role in this bunch-to-bunch vertical dipole coupling.

This technique has the advantage of avoiding coupling from preceding
bunches to the bunch being studied, while also providing information
about the coupling of the motion of one bunch to later bunches via
the EC.  The excitation level can, in principle, be tailored for the
bunch that is being driven; the ability to keep a relatively fixed
oscillation amplitude of the driven bunch could be important for
conditions when the first bunches in the train are more stable but
the latter bunches are not.  This method has the drawback that it is
slower than the preceding method, as it requires collecting
turn-by-turn position data for every bunch times the number of
bunches within the train. It is, therefore, sensitive to drifts in
the tunes of the storage ring.  The data acquisition software is
sufficiently flexible to permit the excitation of the first bunch in
the train (which generally has quite weak coupling due to the EC to
all of the other bunches in the train) in addition to the bunch
actually being studied.  In this way the tunes of the later bunches
are measured with respect to the tune of the first bunch, which
allows for correction of drifts in the tune during the measurement.
Another feature is the ability to turn off feedback for the bunches
which are being excited.  This permits the train of bunches to be
stabilized with feedback, while allowing the bunches, which are
being measured, to have longer damping times, permitting a more
accurate tune measurements and the ability to measure the variation
of damping rates of different bunches within the train.

\subsubsection{Feedback System Response}
\label{sssec:ec_dynamics.hardware_techniques.tunes.fsr}

Another approach for tune measurements became apparent after the
installation of the Dimtel\cite{DIMTEL2009:IGP1281F:Man17} feedback
electronics, capable of damping bunches with spacings down to 4~ns.
While looking at the FFT of the position for a single bunch as part
of the feedback system diagnostics, it was observed that the signal
amplitude varied as a function of the feedback gain. At low gains
the betatron peak is visible, but as the gain is increased the
amplitude of the peak decreases until it become a notch in the
spectrum at high gain.  The notch is created by the feedback system,
whose phase is adjusted to suppress the broadband excitation of the
beam preferentially at the betatron frequency. When the feedback
settings have been fully optimized, the notch in the spectrum for
each bunch marks the location of its betatron oscillation frequency.

\begin{figure}[htbp] 
   \centering
   \includegraphics[width=0.6\columnwidth]{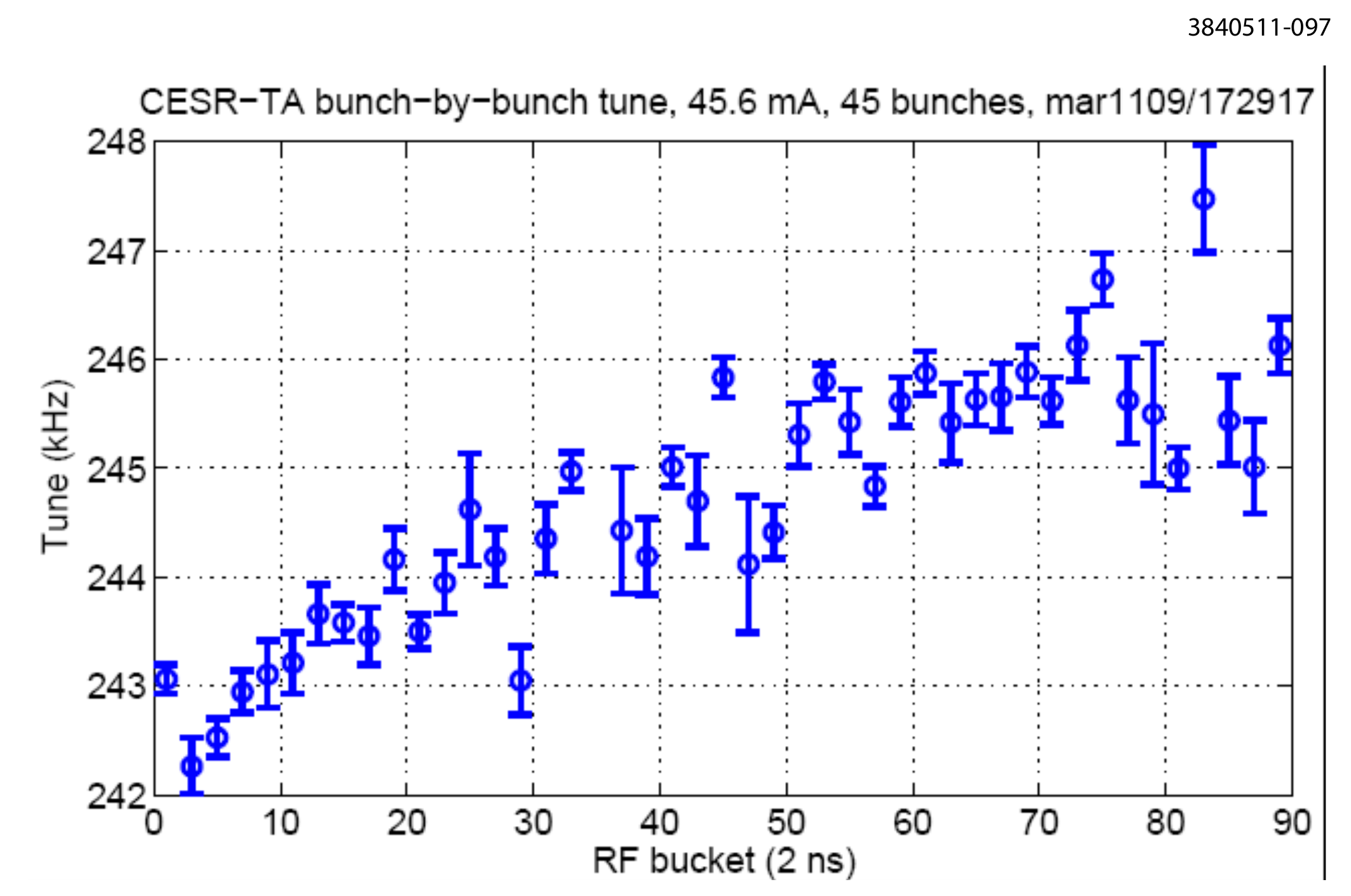}
   \caption[Tune shift measured by 4~ns feedback system]{\label{fig:DimtelTuneShifts}
   Vertical tune vs. RF bucket number for a train of 45~bunches with 4~ns bunch spacing determined from notches in the spectra from the feedback error signal. }
\end{figure}

The position data generally represents the effect of probing the EC
in a regime when the bunches are moving at small amplitudes.  An
example of data taken using this method is seen in
Fig.~\ref{fig:DimtelTuneShifts}.  There is a very clear trend for
the vertical focusing effect from the accumulating EC, which is
visible the plot.  Although this method is quite appealing, only a
few tune shift measurements have been performed via this method.
This technique works well for 4~ns-spaced bunches, but it requires
fairly exact adjustments of the feedback system parameters to be
able to clearly identify the notches in the bunch spectra.  To
obtain the most accurate spectra, the data for each bunch is
averaged typically for 30 seconds, allowing some variation in the
tunes due to longer-term drifts in the storage ring focusing.

\subsubsection{Self-Excitation}
\label{sssec:ec_dynamics.hardware_techniques.tunes.se}

The last method utilized for bunch-by-bunch tune shift measurements
is a by-product of the observation of beam instabilities, described
in the next section.  In this set of measurements the position
spectrum of each bunch is measured with a gated spectrum analyzer.
Two of the peaks that are visible in these self-excited spectra are
the horizontal and vertical dipole modes.  The shift of the tunes as
the gate is moved from bunch to bunch are easily detected via this
method.  Since most of these measurements are taken in conditions
when the beam is above or near an instability threshold for at least
some of the bunches within the train, the self-excited amplitudes of
the dipole motion will vary along the train.  This method is quite
sensitive to low signal levels, with the noise floor for small
amplitude oscillations at the level of 0.4~$\mu$m-rms horizontally
and 0.2~$\mu$m-rms vertically.  Due to averaging in the spectrum
analyzer, the data acquisition requires about 1~minute for each
bunch, which is long enough to make this method sensitive to drifts
in the storage ring tunes.

\subsection{Instability Measurements}
\label{ssec:ec_dynamics.hardware_techniques.instability}

An important set of {CesrTA} measurements focuses on beam
instabilities due to the EC.  These studies measure the growth of
self-excited oscillation amplitudes of the bunch centroids and the
growth of vertical beam size along the train under various
accelerator and electron cloud conditions.  The first piece of
hardware utilized for these measurements is a monitor for the
bunch-by-bunch beam position.  The other detection system required
is the xBSM monitor for determining the vertical beam size of each
bunch.

\subsubsection{Bunch-by-Bunch Position Spectra}

For instability studies, the bunch-by-bunch position measurements
are accomplished by a BPM detector connected to one of CESR's
original relay-based BPM system processors, which in turn passes its
video output signal to a spectrum analyzer in the control room. (See
reference \cite{JINST12:T11006} for the description of the
hardware.)  BPM33W, which is located at a high vertical beta point,
has generally been used as the detector for these observations.  The
signal is taken from one button, making it sensitive to both the
horizontal and vertical motion.  The data-taking software sets the
trigger delay for the sampling gate to select a particular bunch
within the train. For almost all of the data, an RG-174 coaxial
cable is placed within the signal path to limit the bandwidth of the
button signal (giving an effective 20~dB of signal attenuation) and
to this an additional 12~dB of amplification is added.  The signal
is then sent to the biased peak rectifier circuit, which has an
effective bandwidth of 700~MHz, and a decay time constant of
approximately 5~$\mu$s.  The resulting video signal is buffered and
sent on a wide-band coaxial cable to a spectrum analyzer in the
control room.

The spectrum analyzer is a Hewlett Packard model 3588A, operating in
the baseband (in these studies the center frequency ranges from
190~kHz to 310~kHz) in ``Narrowband Zoom'' mode with a 40~kHz span.
This mode of operation performs a $\pm20$~kHz FFT on time slices of
the signal and these spectra are averaged for 100~time slices,
taking about 10 seconds for each 40~kHz step of the center
frequency.  At 2.1~GeV the position sensitivity of the signal from
the BPM at 33W was measured to be

\begin{equation}
\begin{array}{l}
x_{\textrm{rms}}=x_{0}/I_{b}(\textrm{mA})\times10^{A_{x}(\textrm{dBm})/20} \\
\\
y_{\textrm{rms}}=y_{0}/I_{b}(\textrm{mA})\times10^{A_{y}(\textrm{dBm})/20}
 \end{array}
\end{equation}

\noindent where $x_{0}=81.3$~mm and $y_{0}=45.3$~mm, when the RMS
bunch length was approximately 10~mm.  $A_x$ and $A_y$ are the
amplitudes measured on the spectrum analyzer in dBm.  With this gain
configuration and over the frequency range of study, the noise
baseline falls from $-95$~dBm to $-105$~dBm (corresponding in the
vertical direction, respectively, to displacements of 1.1~$\mu$m-RMS
to 0.33~$\mu$m-RMS for a 1~mA~bunch.)

\begin{figure}[htbp] 
   \centering
   \includegraphics[trim= 0mm 80mm 0mm 0mm, clip=true, width=0.5\columnwidth]{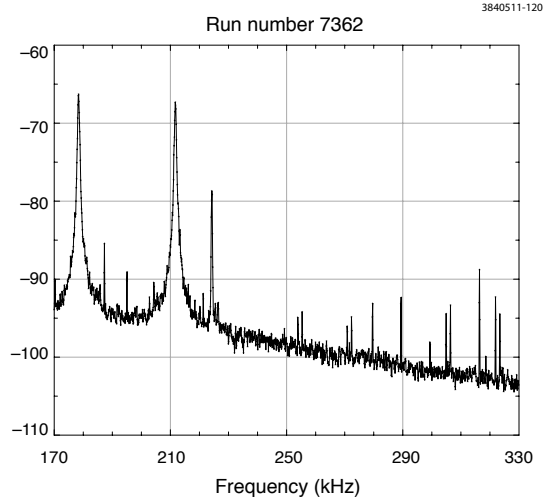}
   \caption[Single bunch EC instability spectrum for bunch~1]{\label{fig:SingleBunchSpectrum-b1}
   Self-excited beam power spectrum for bunch~1 in a 30~bunch-long positron train at 2.1~GeV beam energy. Since this spectrum overlaps the 1/2 integer resonance at
      195 KHz, this frequency is a reflection point for the spectra.  Thus
      the peak at 178 KHz  is a reflection of the peak at 212 KHz. }
\end{figure}

\begin{figure}[htbp] 
   \centering
   \includegraphics[trim = 0mm 80mm 0mm 0mm, clip=true, width=0.5\columnwidth]{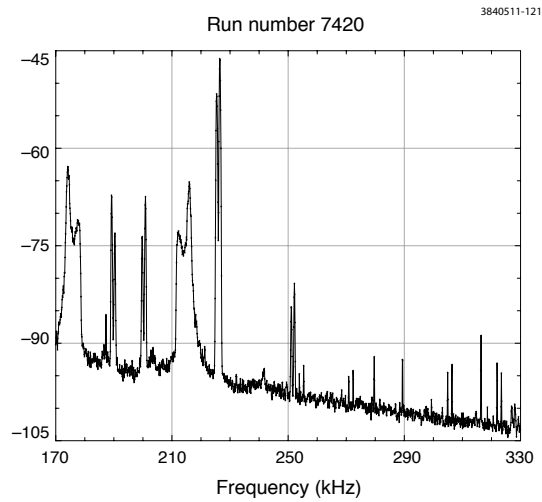}
   \caption[Single bunch EC instability spectrum for bunch~30]{\label{fig:SingleBunchSpectrum-b30}
   Self-excited beam power spectrum for bunch~30 in a 30~bunch-long positron train at 2.1~GeV beam energy. }
\end{figure}

\begin{figure}[htbp] 
   \centering
   \includegraphics[height=0.4\textheight]{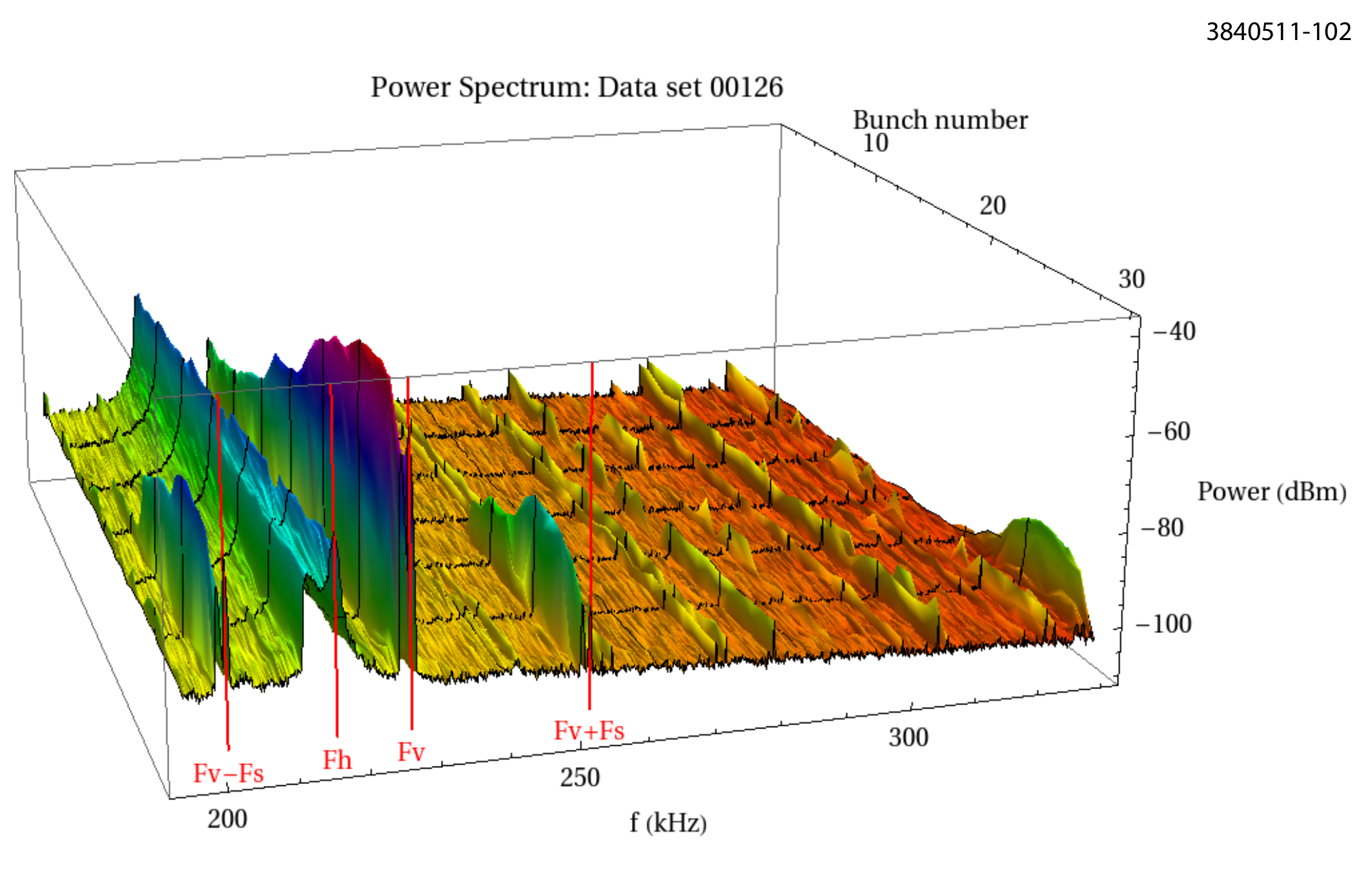}
   \caption[Example of EC instability spectrum for 30~bunch train]{\label{fig:ExampleMultiBunchSpectrum-30Bunches}
   Self-excited beam power spectra for bunches~1 through~30 in a 30~bunch-long positron train at 2.1~GeV beam energy.
   The horizontal axis is the frequency, the vertical axis is the spectral power in dB and the axis into the page is the
   bunch number with bunch~30 being in the foreground.  Red vertical lines in the foreground denote in ascending order the
   location of the $m=-1$ vertical head-tail line, the horizontal tune, the vertical tune and the $m=+1$ vertical head tail line.}
\end{figure}

Representative self-excited spectra of the first and last bunch in a
30-bunch positron train at 2.1~GeV are shown in
Figures~\ref{fig:SingleBunchSpectrum-b1}
and~\ref{fig:SingleBunchSpectrum-b30}.  For this train the
horizontal tunes are in the range from 212~kHz to 218~kHz, and the
vertical tunes are in the range from 224~kHz to 227~kHz.  Since this
spectrum overlaps the $1/2$ integer resonance at 195~kHz, this
frequency is a reflection point for the spectra.  For bunch~30,
additional lines are visible in the ranges 198-201~kHz and
250-252~kHz; these correspond to vertical head-tail modes as their
frequencies are plus and minus the synchrotron oscillation frequency
added to the vertical tune.  The baseline is seen to be falling as
roughly a $1/f$~noise spectrum.  There are also a number of
unrelated noise lines, scattered throughout the spectra assumed to
be due to ``cultural noise sources.''   A ``mountain-range'' plot of
the spectra of all 30~bunches within a 30~bunch-long train is shown
in Figure~\ref{fig:ExampleMultiBunchSpectrum-30Bunches}.  A cut of
the spectrum has been made at the half integer resonance (195~kHz)
to suppress the ``reflected'' spectral lines.  In this plot the
self-excited vertical tune amplitude begins to grow at approximately
bunch 10 and continues to grow in amplitude until near bunch~20.  In
this region the two vertical head-tail lines appear above the noise
background.  Also around bunch~15 the spectral peak of the
horizontal tune appears to bifurcate, something which is also seen
in Figure~\ref{fig:SingleBunchSpectrum-b30}, and on close
examination these data also show bifurcation of the vertical tune
and the vertical head-tail lines for the last bunches in the train.
Figure~\ref{fig:ExampleMultiBunchSpectrum-30Bunches} also shows a
number of ``fences'', i.e. peaks in the spectrum at fixed
frequencies due to external ``cultural noise sources.''

Many tests have examined the self-consistency and interpretation of
the data.  The identification of the vertical and horizontal tunes
was checked by changing the controls for each separately and
verifying which spectral peak moved.  They were also checked using
BPMs at other locations, which had buttons summed to produce
dominantly horizontally- or vertically-sensitive detectors.  The
interpretation that the vertical head-tail lines were not
inter-modulation distortion components coming from the processing
electronics was tested by switching a 6~dB attenuator into the
signal path upstream of the peak detector and observing the change
in both horizontal and vertical spectral peaks. If the head-tail
lines were actually inter-modulation cross-products from the
non-linearity of the electronic processing, then they would have
decreased by 12~dB, and they only decreased by $6\pm2$~dB.

Although this method for detecting the frequency spectra of the
bunches is fairly sensitive, the measurements must be made
separately for each individual bunch.  The measurement time is about
1~minute per bunch for the selected frequency range.  This means
that the data represents a time-average of any unstable motion over
this period. In addition, due to the finite beam lifetime, the beam
must be refilled a number of times during data-taking  for one set
of conditions.  In our case, we choose typically to refill after
measuring spectra for five bunches. When these spectra are plotted,
the beam intensity decay over five bunches gives the amplitude for
the peaks within the spectrum a slightly scalloped shape.  This
refilling cycle coordinates fairly well with the cycle to measure
and readout the bunch-by-bunch and turn-by-turn xBSM data.

It is possible to compare the above measurements to those read out
turn-by-turn and bunch-by-bunch from a number of BPMs via the CBPM
system (which has  a much faster data acquisition time).
Unfortunately the head-tail lines are not visible above the noise
floor in the CBPM data.  Our explanation is that the relay BPM
system peak rectifies the position signal and, if there is a
temporal variation due to head-tail motion, the arrival time of the
signal varies correspondingly.  This gives a frequency modulation to
the position signal when viewed by the spectrum analyzer.  The CBPM
processing is different: the signal is sampled at a fixed time
corresponding to the positive peak of the button BPM pulse.  Any
variation in the arrival time produces only a second order variation
in amplitude and, even if one moved the sampling time significantly
off of the peak, it  does not produce any observable signal at the
head-tail line frequencies.

\subsubsection{Bunch-by-Bunch Beam Size}
\label{sssec:ec_dynamics.hardware_techniques.xray}

At {CesrTA}, bunch-by-bunch beam sizes are measured using an x-ray
monitor \cite{NIMA798:127to134} built on the D Line of the CHESS
light source for viewing positrons.  (A similar line for viewing
electrons is installed at the C Line, useful particularly for
comparison measurements for electron beams vs. positron beams.)  The
detector can read out bunch-by-bunch, turn-by-turn signals at 14~ns
or 4~ns spacing.  Three sets of x-ray optics can be selected in the
optics box:  Coded Aperture (CA), Fresnel Zone Plate (FZP) and an
adjustable slit.  The coded aperture mask permits single-shot,
photon-statistic-limited resolutions of $\sim2-3$ $\mu$m at vertical
beam sizes of less than 20~$\mu$m\cite{DIPAC11:WEOB03},
\cite{NIMA767:467to474}.


During a given set of instability measurements, typically xBSM data
were taken using two sets of optics, the adjustable slit and the CA.
This allows the greatest range of sensitivity for measurements of
the vertical size and centroid motion of the beam.  During the
measurement cycle, the beam size data are taken bunch-by-bunch and
turn-by-turn generally immediately after the train has been topped
off, usually occuring after taking the frequency spectrum for every
fifth bunch.

\subsection{Mode Growth Rates}
\label{ssec:ec_dynamics.hardware_techniques.mode_growth}

A complement to the instability measurements, described in the
preceding section, are the damping rate measurements for the
coherent transverse modes.  The instability measurements easily
record the large amplitude coherent signals as the bunches become
unstable and ultimately limit due to non-linearities in the bunch
dynamics.  However, the damping measurements give information about
the stability of the bunch at small amplitudes before the bunch goes
unstable, the regime in which storage rings and damping rings will
actually operate.  These studies give some insight about how the
beam instability begins developing from earlier to later bunches
along the train.

\subsubsection{Drive-Damp Excitation: Method}

The basic idea for these observations is to employ the same relay
BPM configuration as is used for the instability measurements.
However, the spectrum analyzer's center frequency is adjusted to be
at either the vertical betatron dipole-mode frequency or one of the
vertical head-tail mode frequencies while the spectrum analyzer is
configured to be in ``Zero Span'' mode.  In this mode the analyzer
functions as a tuned receiver with its display producing signal
amplitude vs. time.  The spectrum analyzer's tracking generator
output is connected to the vertical feedback system's external
modulation input.  Aside from the spectrum analyzer's control
settings, this is quite similar to the hardware configuration shown
in Fig.~\ref{fig:SingleBunchDriverDiagram}.  By adjusting the
digital timing controls for the feedback modulator's external input,
it possible to drive only one bunch, as long as the bunch spacing is
greater than 6~ns (if the bunch spacing is 4~ns, then the duration
of the pulse on the beam stabilizing feedback system stripline
kicker is long enough to deflect the bunch under study and also
slightly deflect the following bunch).  To permit the drive-damp
modulation of the beam, there is one additional element added to the
block diagram of Fig.~\ref{fig:SingleBunchDriverDiagram}.  This
element is a modulating gate for the spectrum analyzer's tracking
generator signal.  The modulator gate is timed with the spectrum
analyzer's timing sweep to pass the tracking generator output for
3~ms at the beginning of the sweep and then to gate off its output
until the start of the next sweep.

\begin{figure}[htbp] 
   \centering
   \includegraphics[width=0.4\columnwidth]{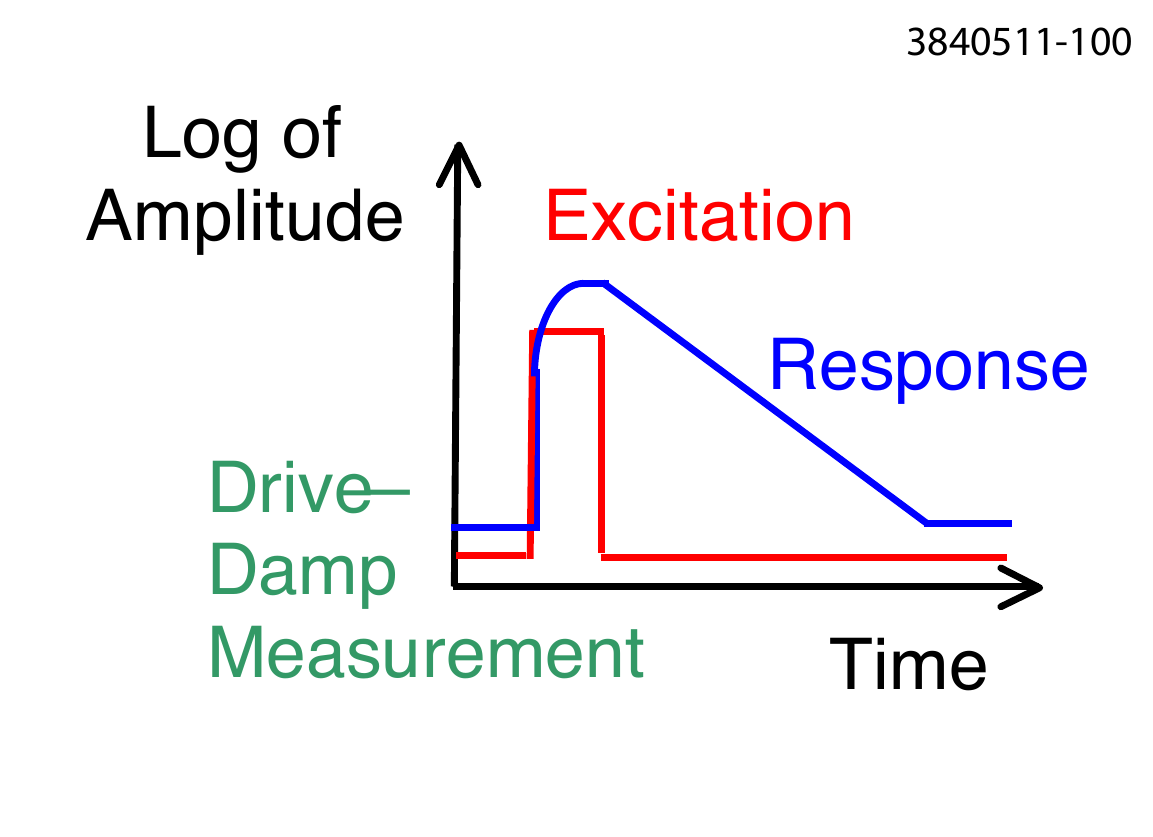}
   \caption[Illustration of drive-damp measurement technique]{\label{fig:DriveDampMeasurement}
   Illustration of the drive-damp measurement:  The red trace is the amplitude of the excitation driving the bunch.  The blue trace is the bunch response. }
\end{figure}

An illustration of the timing and the expected signal response are
shown in Fig.~\ref{fig:DriveDampMeasurement}.  The red curve shows
that the amplitude of the transverse excitation of the bunch vs.
time is an impulse.  The expected beam response initially grows
during the driving impulse, usually reaching a saturated level, and
then decays exponentially after the drive is switched off (shown in
the logarithmic plot as a linear decrease vs. time).  If the
frequency of the spectrum analyzer tracking generator is tuned away
from the bunch resonant frequency, the decaying response can have
more than one frequency component, resulting in periodic oscillatory
beats.  So during the measurement it is necessary to make small
tuning adjustments to the excitation frequency to produce the most
exponential decay possible.

The excitation of the bunch is accomplished in a somewhat different
manner for the betatron dipole mode and the head-tail modes.  In
both cases the frequency of the spectrum analyzer is set to drive
the coherent mode frequency being measured.  However, for the
head-tail modes it is necessary to also continuously drive the
external modulation input for RF cavity phase at the synchrotron
oscillation frequency.  A basic explanation is provided here, but
more detailed calculations and simulations may be found in a
subsequent paper. This imposes a longitudinal energy oscillation on
all of the bunches within the train, causing them to uniformly shift
their arrival times and displace the train centroid horizontally
proportional to the local dispersion.  The typical amplitude of this
oscillation is relatively large, with the peak fractional energy
varying as much as $\pm7.6\times10^{-3}$ for all of the bunches
within the train.  Due to the RF system's non-linearities, there may
be some increase in the energy spread (and bunch length) of the
bunches. In the presence of the large energy variation, the
transverse field from the stripline kicker deflects the lower energy
particles in the bunch (displaced toward the head of the bunch) more
than the higher energy particles (displaced toward the tail of the
bunch.) Although this is a fairly small differential effect, the
bunch is being driven on the head-tail resonance, allowing the
head-tail coherent oscillation amplitude to grow.

\subsubsection{Drive-Damp Excitation: Examples}

\begin{figure}[htbp] 
   \centering
   \includegraphics[width=0.6\columnwidth]{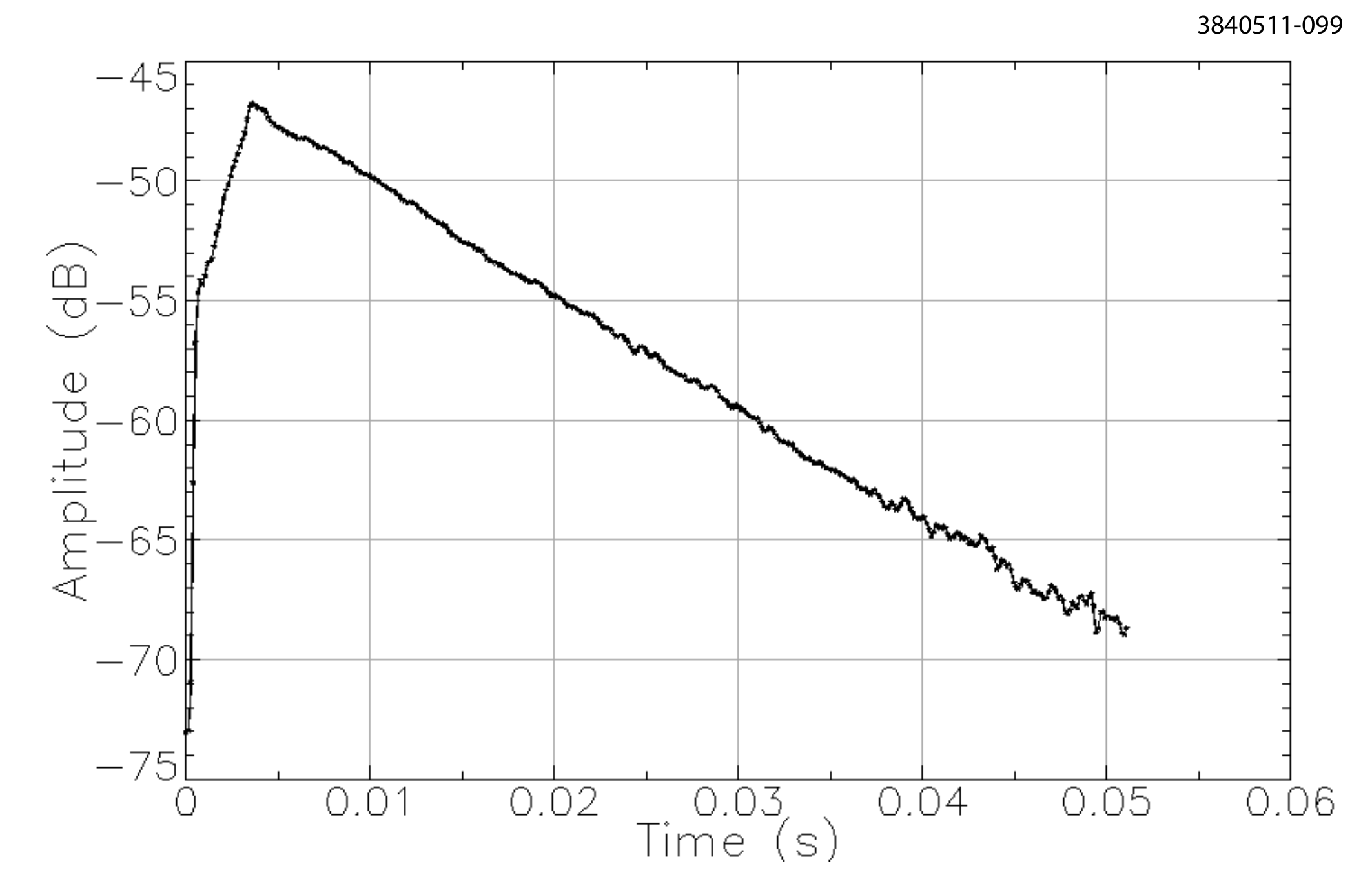}
   \caption[Drive-damp measurement for dipole betatron mode]{\label{fig:DriveDamp029808}
   Drive-damp measurement:  The trace is the response for the bunch being driven at the vertical betatron frequency.
   The vertical and horizontal scales are 5~dB and 10~ms per division, respectively. }
\end{figure}

\begin{figure}[htbp] 
   \centering
   \includegraphics[width=0.6\columnwidth]{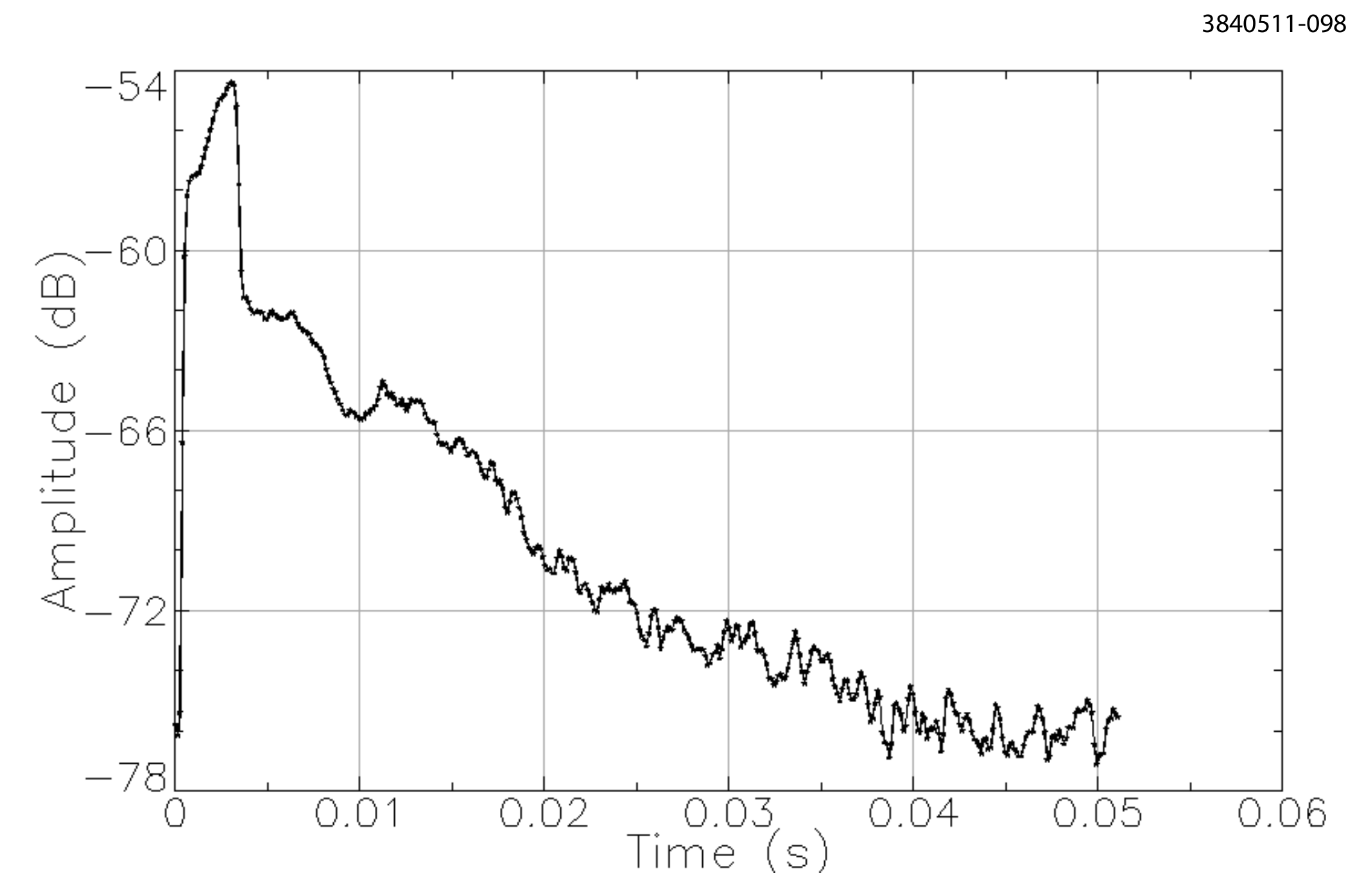}
   \caption[Drive-damp measurement for head-tail mode]{\label{fig:DriveDamp007420}
   Drive-damp measurement:  The trace is the response when one of the head-tail modes is excited.
   The vertical and horizontal scales are 6~dB and 10~ms per division, respectively. }
\end{figure}

Two examples of actual drive-damp measurements are found in
Figs.~\ref{fig:DriveDamp029808} and~\ref{fig:DriveDamp007420}.  In
Fig.~\ref{fig:DriveDamp029808} the betatron dipole mode amplitude
ramps up for the first 3~ms and then decays exponentially
thereafter.  Fig.~\ref{fig:DriveDamp007420} shows one of the
head-tail modes being excited.  The initial 7~dB drop in the
amplitude of the signal represents the off-resonance excitation of
the dipole mode, which immediately switches to oscillating at its
resonant frequency (outside of the bandwidth of the receiver) when
the drive turns off; the roughly exponential shape thereafter is the
head-tail mode decay.  As a test, the longitudinal drive to the
phase of the RF cavity was turned off, and the head-tail mode
exponentially damped signal was observed to vanish into the noise
floor.

This type of measurement may be very useful for understanding the
behavior of bunches within the train before their motion becomes
unstable.  However, even though much of the data acquisition is
automated, there are a few steps in the present version of the data
acquisition software, which must be completed manually.  In
particular the fine adjustment of the spectrum analyzer frequency
(centering it on the coherent mode frequency) is necessary to
optimize the exponential damping curve.  The manual adjustment of
the frequency makes this type of measurement fairly time-consuming.
Routinely, after data are taken for several bunches, the beam is
topped off.  Beam size measurements are typically taken immediately
after topping off.


\section{Summary}
\label{sec:summary}

This paper has described the instrumentation, which has been
developed or modified for use in the CesrTA~ program for the
investigation of storage ring beam dynamics.  It also has provided
representative examples of actual data and explanations how the
basic analysis of some of the data is accomplished.  In particular
these studies have focused on the methods for low emittance tuning
of the beam, on the causes for intra-beam scattering of single
bunches and for production and interaction of bunches within trains
with electron clouds, produced by photo-electrons from synchrotron
radiation and secondary emission.

\bibliographystyle{JHEP}
\bibliography{CesrTA}

\end{document}